\documentclass[preprint,showpacs,nofootinbib,amsmath,amssymb,floatfix]{revtex4}

\usepackage{epsfig}
\usepackage{bm}

\begin{document}

\bibliographystyle{apsrev}
\title{Oscillation Effects and Time Variation of the Supernova Neutrino Signal}

\author{James P. Kneller}
\surname{Kneller}
\email{kneller@physics.umn.edu} 
\affiliation{School of Physics and Astronomy, University of Minnesota, Minneapolis, Minnesota 55455}

\author{Gail C. McLaughlin}
\surname{McLaughlin}
\email{Gail_McLaughlin@ncsu.edu} 
\affiliation{Department of Physics, North Carolina State University, Raleigh, North Carolina 27695-8202}

\author{Justin Brockman \footnote{Current address: Department of Applied Physics, Stanford University, Stanford, CA 94305-4090} }
\surname{Brockman}
\email{Justin_Brockman@ncsu.edu} 
\affiliation{Department of Physics, North Carolina State University, Raleigh, North Carolina 27695-8202}

\date{\today}




\begin{abstract}

The neutrinos detected from the next Galactic core-collapse supernova will contain 
valuable information on the internal dynamics of the explosion. 
One mechanism leading to a temporal evolution of the
neutrino signal is the variation of the induced neutrino flavor mixing 
driven by changes in the density profile.
With one and two dimensional hydrodynamical simulations we identify 
the behavior and properties of prominent features of the explosion. Using
these results we demonstrate the time variation of the neutrino crossing probabilities 
due to changes in the MSW neutrino transformations as the star explodes 
by using the S-matrix - Monte Carlo - approach to
neutrino propagation. After adopting spectra for the neutrinos emitted from the 
proto-neutron star we calculate for a Galactic supernova the evolution of the positron spectra 
within a water Cerenkov detector and the ratio of charged current to neutral current event rates 
for a heavy water - SNO like - detector and find that these detector 
signals are feasible probes of a number of explosion features.

\end{abstract}

\maketitle

\newpage

\section{Introduction} \label{sec:intro}

The importance of neutrinos in the explosions of massive stars has 
long been recognized. This significance is coupled with their ability to 
carry information to us about the processes and conditions in the core of 
the supernova so that the neutrino signal from the next Galactic supernova will 
provide us with an opportunity to test the core collapse paradigm. 
The explosion begins when the runaway process of 
electron capture in the core a massive star leads to a rapid 
compression and collapse that is only halted when the 
degenerate pressure of the neutrons and thermal pressure of non-degenerate 
particles kicks in at super-nuclear densities. The rapid
neutronization during the collapse leads to a large burst of $\nu_{e}$ neutrinos that 
identifies the beginning of the event. As nuclear densities are reached 
the mean free path of the neutrinos becomes shorter than the size of the 
proto-neutron star and the neutrinos become trapped. Thermal processes 
within the core create a thermal bath of 
neutrino-antineutrino pairs of all flavors. The neutrinos slowly diffuse 
from the core over a period of order $10$ seconds, carrying 99\% of the gravitational
binding energy of the core with them. With such a large neutrino 
luminosity even a small number of neutrino 
interactions above the core can create an important impact on the explosion physics.
Detection of the neutrinos from the next nearby supernova will offer us the opportunity 
to examine the internal evolution of the explosion 
since the neutrinos allow us to see down all the way 
to the proto-neutron star. For example,
if the proto-neutron star collapses to a black hole, the resulting neutrino signal
will be altered \cite{Beacom:2000qy,McLaughlin:2006yy}. 

In addition, although these neutrinos are only weakly interacting they can have significant
effects upon the nucleosynthesis that occurs in the supernova:
in the outer layers of the star neutrinos can cause 
a transformation of the elements synthesized during the preceding thermonuclear 
burning period in the ``neutrino process'' \cite{Hartmann:1991tk,Heger:2003mm}, 
it is thought that a neutrino driven wind occurs at late time in the supernova which 
may create sufficiently neutron rich material 
to produce the r-process elements \cite{Woosley:1994ux}, and 
long duration gamma ray bursts (which are thought to be a rare type of supernova)
produce elements in conditions governed by strong neutrino fluxes
\cite{Surman:2005kf,Pruet:2003yn}. 
The results of all these nucleosynthesis processes are quite 
sensitive to the neutrino spectra 
\cite{McLaughlin:1997qi,McLaughlin:1996eq, Meyer:1998sn,Yoshida:2006sk}. 

For reasons pertaining to both hydrodynamics and to element synthesis 
it is important to understand the details of neutrino propagation in the supernova and the mixing between 
neutrino flavors. Neutrino oscillations also alter 
what we observe but, simultaneously, also allow the possibly of elucidating 
yet unknown neutrino oscillation physics, e.g. 
\cite{FHM1999,Dighe:1999bi,Engel:2002hg,lunardini,Friedland:2003dv}.
Supernova neutrino flavor transformation is a rapidly
developing field with the neutrino 
background terms the subject of intense study 
\cite{Duan:2006jv,Sawyer:2005jk,Hannestad:2006nj}.
Realistic possibilities include a complete mixing of all flavors \cite{Sawyer:2005jk} or a 
partial oscillation between flavors \cite{Duan:2006jv}.
An important part of any future observation of supernova neutrinos will be to 
understand this physics.
In parallel, phase effects due to multiple resonances have recently been recognized 
\cite{Kneller:2005hf,Dasgupta:2005wn} distorting the simple picture of neutrino propagation 
through the outer layers of the supernova that has been used 
in the past \cite{FHM1999,Dighe:1999bi,Engel:2002hg}. 
The explosion of the star leads to an evolution of the flavor mixing so 
that to appreciate the effects upon neutrino propagation
it is important to understand the density profile. 

Observations of high pulsar velocities and polarized supernova light 
suggest an inherent asymmetry in the explosion mechanism. Though many 
possible mechanisms for generating the asymmetry have been proposed, 
recent work by Blondin, Mezzacappa \& DeMarino \cite{2003ApJ...584..971B} has identified an instability of the 
standing accretion shock that leads to large dipole and quadrupole 
moments. More recent work by Blondin \& Mezzacappa \cite{BM2006} and Ohnishi, Kotake, 
and Yamada \cite{OKY2006} as well as Scheck \emph{et al.} \cite{SKJM2006} 
has both confirmed and furthered the understanding of this instability. 
An alternative mechanism, acoustic heating, has also been 
discussed by Burrows et al. \cite{Burrows:2006uh}. The temporal evolution of 
the neutrino flavor mixing and the subsequent variation of the neutrino signal 
means that it may be possible to to detect prominent 
features of the supernova density profile and to use this information to
learn important information about the explosion.
This possibility was demonstrated first by Schirato and Fuller \cite{SF2002} 
who used a time-dependent, one-dimensional $1 / r^{2.4}$ density 
profile to demonstrate that the supernova's forward shock wave reaches 
and disrupts the `H' resonance transformation layer a few seconds after 
the core bounce leading to a detectable change in the charged current neutrino signal. This was 
followed and elaborated upon by Dighe \& Smirnov \cite{Dighe:1999bi}, Takahashi 
\emph{et al.} \cite{TSDW2003} and Fogli \emph{et al.} \cite{Fetal2003}. 
More recently Tom{\` a}s \emph{et al.} \cite{2004JCAP...09..015T} used 
a more sophisticated numerical model of a progenitor star to investigate 
the effect of both forward and reverse shocks on the neutrino signal.
Tom{\` a}s \emph{et al.} also present one second of a two-dimensional 
simulation result to show how the density profile can be greatly complicated
by a deformed forward shock followed by strong convection currents. 
But due to the intensive computational burden of supernova models 
it is numerically very taxing to to watch these the explosion propagate 
outward until they affect neutrino transformations. 

In this paper we present calculations which link hydrodynamical simulations of 
time dependent density profiles and phase-retaining neutrino oscillations. 
Our presentation begins in section \ref{section:simulations} with the results 
from simulations of supernovae in both one and two dimensions where we 
artificially heat a density profile constructed so as to 
mimic the state of the supernova at the point where the accretion shock has stalled. 
Our two-dimensional explosions are heated aspherically so as 
to create an aspherical supernova. Though this 2D model lacks some structural features 
present in more sophisticated models that generate asphericity 
via standing accretion shock instabilities, we still obtain results that give a good idea 
of how density isotropy affects neutrino flavor transformations. 
In section III we pass a spectrum of neutrinos through the simulation results 
and show how the crossing probability is affected by the evolution of the 
profiles. We finish in section IV by presenting our calculation of two 
neutrino detector signals - the positron spectra 
within a water Cerenkov detector, and the ratio of charged 
current to neutral current event rates for a heavy water detector - that 
demonstrate our ability to extract information about the supernova.


\section{The SN Profile} \label{section:simulations}

\subsection{The SN Simulations}

Our first task in determining the neutrino signal is to simulate the supernova so as to 
obtain the time evolution of 
the density profile. A supernova simulation is a complex and computationally intensive problem. 
At the present time 
there is not yet a robust, self-consistent model for core-collapse supernovae, and the most 
sophisticated multi-dimensional simulations take a very long time to compute. But we 
do not need to simulate 
the actual core collapse and formation of the proto-neutron star, these events occur 
deep within the core at 
a radius of $10^{7}\;{\rm cm}$ or so, whereas the first neutrino resonance occurs at around $10^{9}\;{\rm cm}$ 
for neutrino 
energies of order $10\;{\rm MeV}$, and the second is even further out. What is occurring 
in the core is 
irrelevant for our purposes, all we need from the simulation of the supernova is the propagation of the shock through
the star. For this reason we employ a simplistic and artificially energetic numerical simulation to
create explosions with a range of features that might be present in reality. We do not intend this
model to be realistic itself, but rather a tool to help with the difficult problem of reconstructing density features
of an exploding star based on changing neutrino signals detected at Earth.

\begin{figure}
\includegraphics[width=5in]
{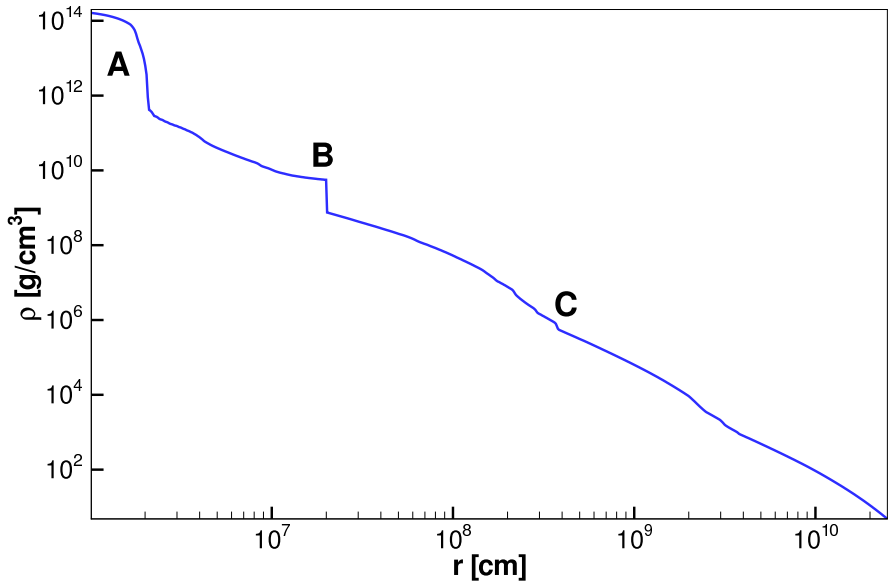} 
\caption{\label{fig:1Da}The initial, $t=0$, density profile used in our SN simulations. The dense core (A) 
inside $20\;{\rm km}$ contains approximately
$3 M_{\odot}$. The slow, outward-moving standing accretion shock (B) is located at $200\;{\rm km}$. 
Above that we have a collapsing, $13.2\;M_{\odot}$, progenitor (C).}
\end{figure}

We use VH-1 \cite{VH-1}, a hydrodynamic code based on Woodward and Colella's \cite{Colella} 
piecewise parabolic method. For both the one-dimensional and two-dimensional simulations we employ an 
exponentially coarsening radial grid that ranges from $r = 10\;{\rm km}$ 
to $r = 250,000\;{\rm km}$ where $r$ is the radius from the center of the star, and 
for the two-dimensional simulations an angular grid that ranges from 
$\theta = 0$ to $\theta = \pi$ where 
$\theta$ is the polar angle. The innermost radial grid increment is set 
to $\delta r = 202.6\;{\rm m}$ and increases by 1.3\% thereafter. By $r = 10^{3}\;{\rm km}$ 
the radial increment has grown to $\delta r \sim 6.5\;{\rm km}$ and by 
$r = 10^{5}\;{\rm km}$ we reach $\delta r \sim 650\;{\rm km}$. 
We map into the code a spherically
symmetric progenitor profile intended to represent the state of the supernova about $100\;{\rm ms}$ after the collapse. 
Beyond $r = 200\;{\rm km}$ the profile is that of a low metallicity, $13.2\;M_{\odot}$ progenitor model 
developed by Heger et al \cite{Heger} that we have allowed to collapse
further within VH-1 so that the infall velocities approached free fall. Interior to $r= 200\;{\rm km}$ we splice 
in a slow, outward-moving standing accretion shock profile provided by Heywood \cite{Heywood} that is similar to that 
found in \cite{2003ApJ...584..971B}. Finally, inside $r=20\;{\rm km}$ we insert a dense, $\gamma = 2.5$, 
polytropic core containing approximately $3 M_{\odot}$. The inner boundary
condition inside the dense core at $10\;{\rm km}$ is reflecting. 
Figure (\ref{fig:1Da}) shows the full initial density profile.
To drive the explosion, we mimic neutrino heating by inserting energy into the region
above a $r_{g}=100\;{\rm km}$ gain radius. The energy deposition rate per unit volume, $dQ/dtdV$, 
is proportional to the density of material, falls as $1/r^{2}$ and 
decreases exponentially with time $t$ over a timescale $\tau$ set to $\tau=0.5\;{\rm s}$. 
For the 1D simulations then,
\begin{equation}
\frac{dQ}{dtdV} \propto \frac{\rho}{r^{2}}\,\left( \frac{r-r_{g}}{r} \right)\,t\,e^{-t/\tau}.
\end{equation}
The additional factors of $(r-r_{g})/r$ and $t$ are inserted so as to ramp up the 
energy deposition over both distance and time thereby avoiding 
the violent disruption that occurs if we deposit energy either too suddenly or within a small volume. 
For the 2D models we introduce an angular dependence into the energy deposition prescription so as 
to match the observation by Blondin \emph{et al.} \cite{2003ApJ...584..971B} that 
small perturbations in standing accretion shock
models can ultimately lead to aspherical shock modes. We break the spherical symmetry of 
our 2D simulations by heating the initial profile with a combination of 75\% 
spherical mode heating and 25\% $\sin^2\theta$ mode heating.
Thus for the 2D cases $dQ/dtdV$ becomes
\begin{equation}
\frac{dQ}{dtdV} \propto \frac{\rho}{r^{2}}\,\left( \frac{r-r_{g}}{r} \right)\,t\,e^{-t/\tau}\,(1+0.5\,sin^{2}\,\theta)
\end{equation}
In either case the total energy, $Q$, input during the simulation was recorded. 
Below the gain radius we wish to maintain the spherical density/gravity conditions and to prevent mass from escaping 
the core so that we do not disturb the shock heating and convective flow in the outer star. 
In a real supernova the core is stabilized by neutrino emission but in lieu of full 
implementation of this cooling mechanism our simulations achieve the same result by forcing all radial and angular 
velocities below $r_{g}$ to zero. In this way we separate the evolution of the outer layers of the star 
from the complex behavior near the core. The runtime of the 1D simulations is 
sufficiently short that we can vary the total energy 
deposition thereby obtaining a variety of results ranging from weak explosions through to the very powerful. 
The 2D simulations take much longer to run and so we have just one data set with an explosion energy set to
$3\times10^{51}\,{\rm erg}$. 

Before presenting our simulation results we mention that 
in general, numerical schemes tend to spread the shock front over several zones and they therefore
become artificially softened.
Schirato and Fuller \cite{SF2002} account for this 
artifact by steepening by hand the density profile at
the shocks. The shocks in our simulations are sufficiently 
steep to demonstrate the primary effect on neutrino mixing, therefore we do not implement a
similar correction.


\subsection{Simulation Results}

The density profiles of the SN we obtain from the simulations possess noticeable differences as the 
energy deposition changes. 
For weak explosions the profiles are monotonically decreasing functions of the radius, $r$, with a 
single, forward facing, 
shock front that moves relatively slowly outwards. As the deposition energy increases a lower-density cavity forms 
behind the shock. Further increases in $Q$ eventually 
lead to the formation of a reverse shock behind the lower-density zone. All three features are present in the 
results of the 2D simulations but the higher dimensionality, coupled with the aspherical heating 
means, that this SN possesses a much more turbulent/chaotic profile. 
\begin{figure}
\includegraphics[width=5in]
{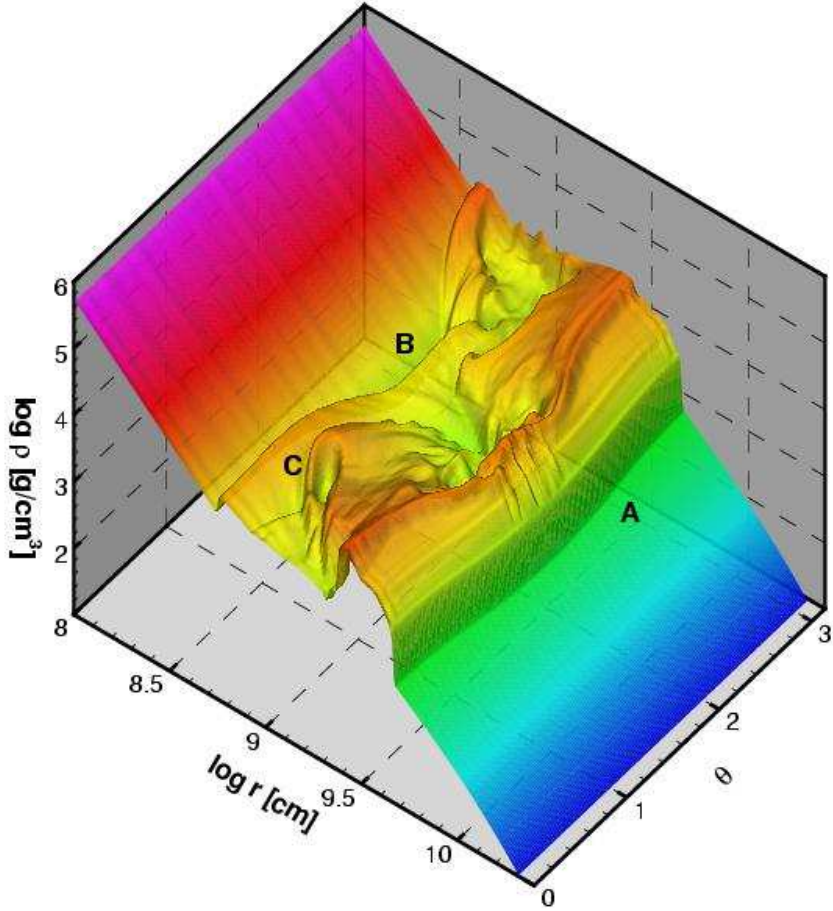}
\caption{\label{fig:rho.snv2d.t=2.5} The density as a function of the radius and angle in a 2D SN model 
at $t=2.5\;{\rm s}$. The forward shock is located to the left of `A', the reverse shock is the step-up in 
density found to the right of `B', and one of the many local cavities in the profile between the shocks 
is to the right of `C'.}
\end{figure}
In figure (\ref{fig:rho.snv2d.t=2.5}) we show a time slice of the 2D SN simulation. The forward shock 
is seen at `A', the `reverse shock' is at `B' and one of the many bubbles in the density profile is located at `C'. 
We note that although our applied neutrino heating is symmetric about the equator ($\theta = \pi/2$) the 
resulting density profile is not. We attribute this to the inherent instability of 
multi-dimensional accretion shocks, potentially excited by simulation numerics. In what follows 
we discuss the behavior and properties of the various features of the profile. 


\subsubsection{The Forward Shock}

The forward shock is a generic feature of supernovae simulations. A stalled forward shock was present in the 
initial profile inserted into the hydrodynamical code and the heating we introduced was meant to revive 
it's outward motion. After revival the shock propagates out through the star and is the explosion feature furthest from
the proto-neutron star. The forward shock is visible in figure (\ref{fig:rho.snv2d.t=2.5}) as the large 
jump in density at larger radii and we note that the forward shock in the 2D simulation is slightly oblate 
due to the aspherical heating of the material. 
We study the behavior of the forward shock with various explosion strengths using the 
1D simulations. 

\begin{figure}
\includegraphics[width=5in]
{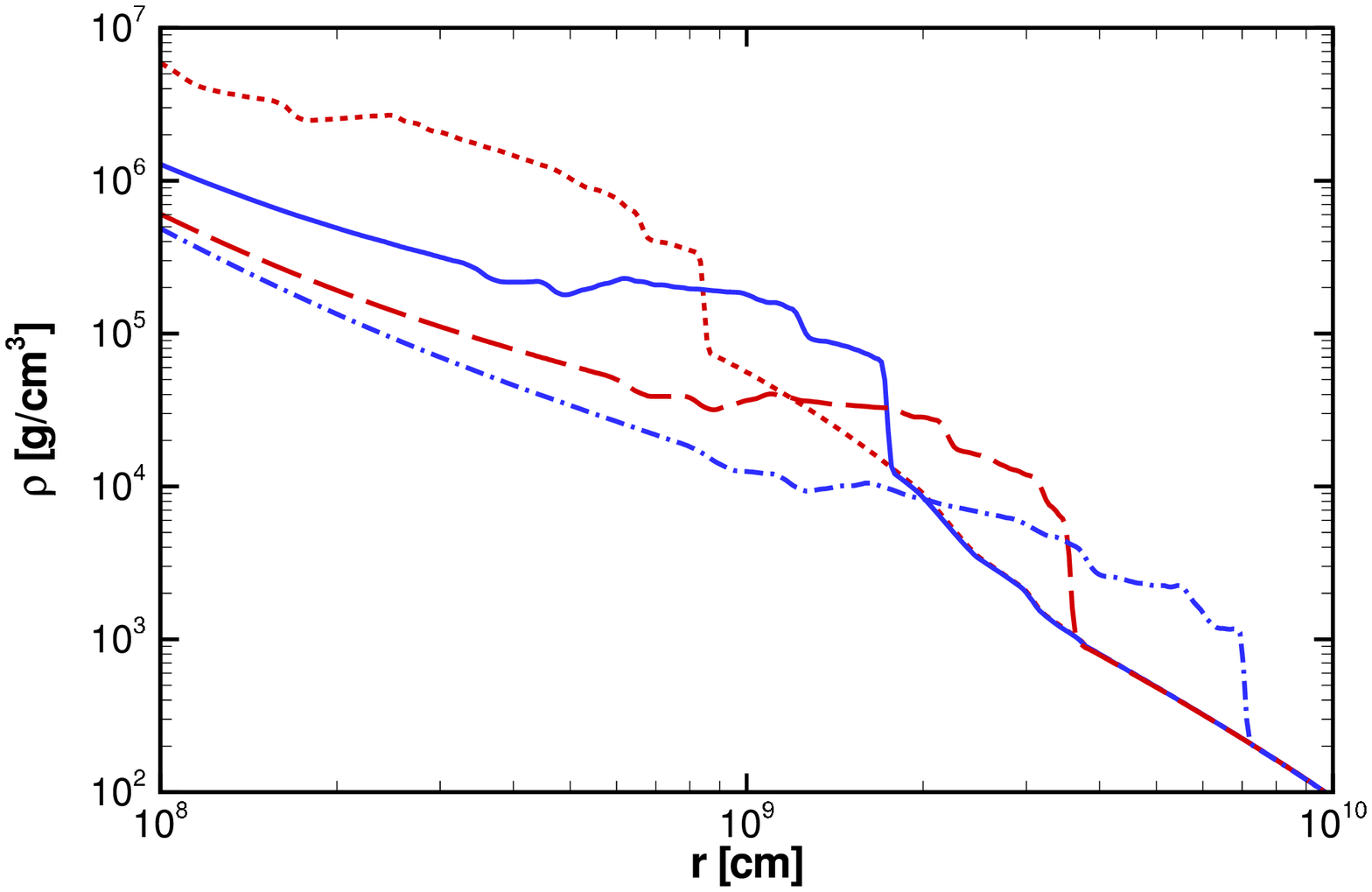}
\caption{\label{fig:rho.snv1d.1e35.t=0.9,1.8,3.6,7.2} The density as a function of the radius in a 1D SN model 
with $Q=1.66 \times 10^{51}\;{\rm erg}$ at $t=0.9\;{\rm s}$ (dotted), $t=1.8\;{\rm s}$ (solid), 
$t=3.6\;{\rm s}$ (long dashed) and $t=7.2\;{\rm s}$ (dash dot).}
\end{figure}
In Fig. (\ref{fig:rho.snv1d.1e35.t=0.9,1.8,3.6,7.2}) we show four snapshots of the density profile in 
a one-dimensional simulation in which the total energy deposition was $Q=1.66 \times 10^{51}\;{\rm erg}$. 
Actual SN are thought to be more energetic than this so we regard the results from 
this simulation as being at the lower end of realistic possibilities. 
This energy is also significantly less than that used in the simulation shown in 
Fig. (\ref{fig:rho.snv2d.t=2.5}) and, consequently, the profile is much simpler. 
The forward shock is clearly visible and we note that the fractional jump in density 
across it does not vary. For normal shocks the density jump across the shock is given by 
\begin{equation}
1+\frac{\Delta \rho}{\rho} = \frac{(\gamma + 1) M^{2} }{(\gamma - 1) M^{2} + 2} \label{eq:delta rho}.
\end{equation}
where $M$ the Mach number and $\gamma$ is determined by the ratio of specific heats. In our 
simulations we adopted a fixed equation of state and the Mach number of the shock is always 
sufficiently large that 
we find that the density jump does not vary to any great extent with time and/or with $Q$. 
But the position of the forward shock as a function of time is greatly influenced by $Q$ with larger 
values leading to more rapidly moving forward shocks. 
\begin{figure}
\includegraphics[width=5in]
{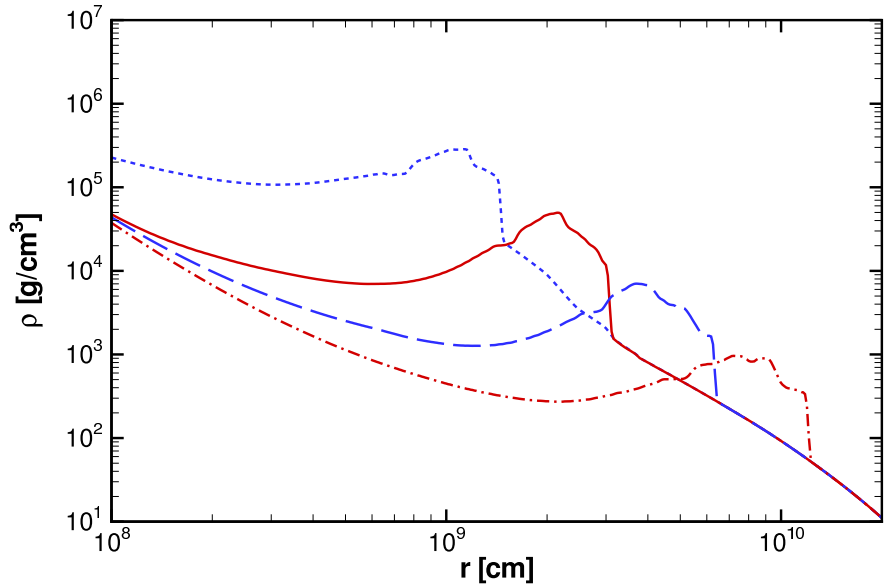} 
\caption{\label{fig:rho.snv1d.6e35.t=0.9,1.8,3.6,7.2} The density as a function of the radius in a 1D SN model 
with $Q=3.07 \times 10^{51}\;{\rm erg}$ at $t=0.9\;{\rm s}$ (dotted), $t=1.8\;{\rm s}$ (solid), 
$t=3.6\;{\rm s}$ (long dashed) and $t=7.2\;{\rm s}$ (dash dot).}
\end{figure}
This can be seen in Fig. (\ref{fig:rho.snv1d.6e35.t=0.9,1.8,3.6,7.2}) which are snapshots at the same 
moments as those in Fig. (\ref{fig:rho.snv1d.1e35.t=0.9,1.8,3.6,7.2}) for a 1D simulation with 
$Q=3.07 \times 10^{51}\;{\rm erg}$. This figure also displays the low-density cavity that can form behind 
the forward shock as $Q$ increases. This profile is very similar to the profiles used by 
Fuller \& Schirato \cite{SF2002} and Fogli \emph{et al.} \cite{Fetal2003}.


\subsubsection{The Reverse Shock}

The heating that led to the regeneration of the forward shock
continues to accelerate the material above the proto-neutron star 
even after the shock has been revived and is moving outwards. A wind is created with a velocity 
that increases with radius. When the velocity of the material becomes larger than the 
local sound speed a reverse shock is formed. 
The reverse shock feature was not present in the `initial' profile: it develops only later. 
If a reverse shock is to form enough energy must be deposited 
to create a sufficiently strong wind. A reverse shock can be seen in figure (\ref{fig:rho.snv2d.t=2.5}) where it
is the step-up in density at the back of the turbulent zone behind the forward shock. 
The reverse shock is oblate due to the aspherical heating 
of the material in the simulation.

\begin{figure}
\includegraphics[width=5in]
{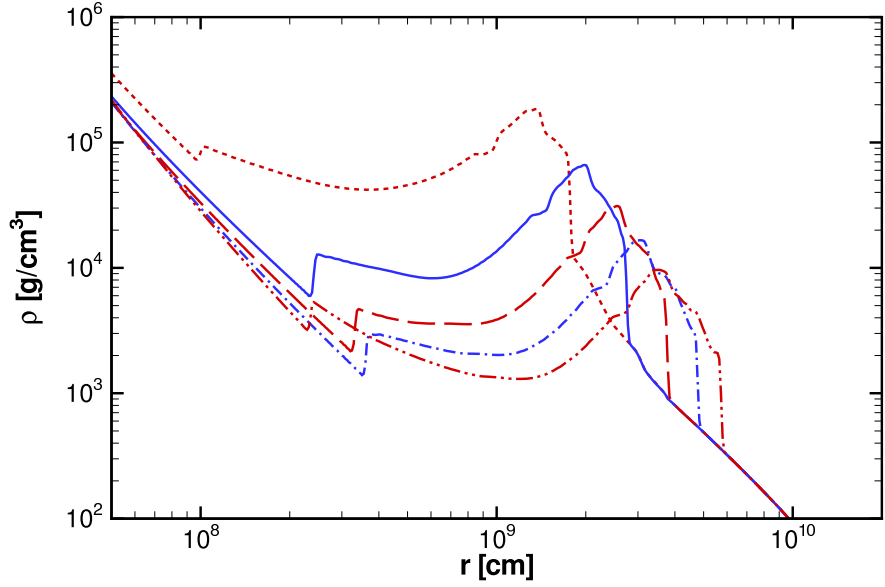}
\caption{\label{fig:rho.snv1d.8e35.t=1,1.5,2,2.5,3} The density as a function of the radius in a 1D SN model 
where $Q=3.36 \times 10^{51}\;{\rm erg}$ at $t=1\;{\rm s}$ (dotted), $t=1.5\;{\rm s}$ (solid), $t=2\;{\rm s}$ (long dashed), 
$t=2.5\;{\rm s}$ (dash-dot) and $t=3\;{\rm s}$ (dash double dot).}
\end{figure}
Once again we can use 1D simulations to study this feature more easily. 
No reverse shock was seen in the profiles shown figure (\ref{fig:rho.snv1d.1e35.t=0.9,1.8,3.6,7.2}) 
or (\ref{fig:rho.snv1d.6e35.t=0.9,1.8,3.6,7.2}) but when we increase $Q$ to
$Q=3.36 \times 10^{51}\;{\rm erg}$ we obtain the results presented in figure 
(\ref{fig:rho.snv1d.8e35.t=1,1.5,2,2.5,3}). 
We notice that the forward shock is, again, 
moving more rapidly compared to the results shown in figures (\ref{fig:rho.snv1d.1e35.t=0.9,1.8,3.6,7.2}) 
and (\ref{fig:rho.snv1d.6e35.t=0.9,1.8,3.6,7.2}) due to the larger energy deposition $Q$. 
The figure also indicates that the reverse shock is smaller 
than the forward shock. 
As $Q$ increases further the density jump across the reverse shock increases 
and it moves closer to the forward shock. We find that it is also possible for the 
reverse shock to penetrate to densities lower than the forward shock. 
\begin{figure}
\includegraphics[width=5in]
{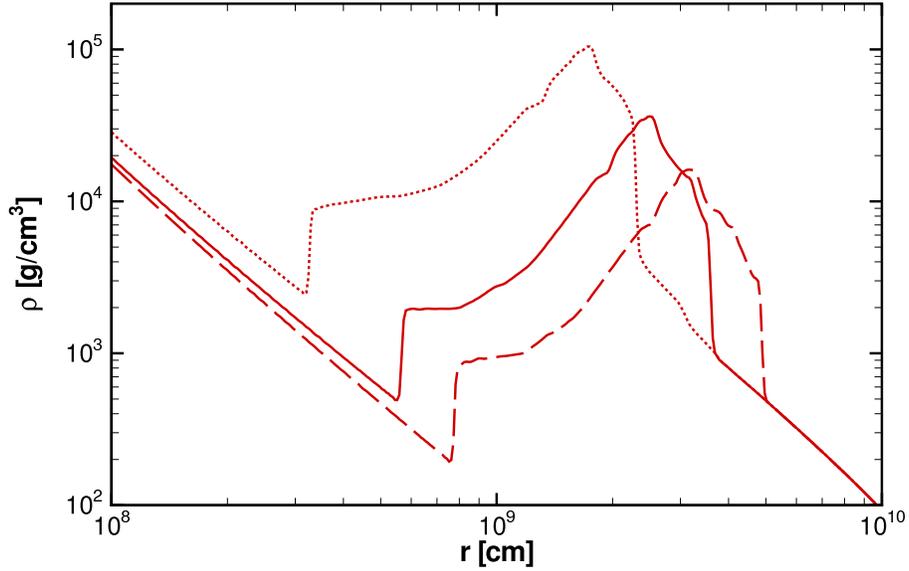}
\caption{\label{fig:rho.snv1d.2e36.t=1,1.5,2} The density as a function of the radius in a 1D SN model
with $Q=4.51 \times 10^{51}\;{\rm erg}$ at $t=1.0\;{\rm s}$ (dashed), $t=1.5\;{\rm s}$ (solid) 
and $t=2.0\;{\rm s}$ (long dashed).}
\end{figure}
Both of these behaviors are shown in figure (\ref{fig:rho.snv1d.2e36.t=1,1.5,2}) 
where we show snapshots of the profile for a 1D simulation with $Q=4.51 \times 10^{51}\;{\rm erg}$. 
The density jump across the 
reverse shock is still smaller than across the forward but they are almost equivalent, and the density 
immediately behind the reverse shock is lower than the density immediately in front of the forward shock. 
It was mentioned by Tom{\`a}s \emph{et al.} \cite{2004JCAP...09..015T} that this feature was also seen at early times 
in their simulations. But realistically SN may struggle to achieve explosions with comparable energies so 
we regard the results of this simulation as being at the upper end of possibilities. 

In both figures (\ref{fig:rho.snv1d.8e35.t=1,1.5,2,2.5,3}) and (\ref{fig:rho.snv1d.2e36.t=1,1.5,2}) the reverse shock was 
driven radially outward by a wind generated by the material heated above the proto-neutron star. Our energy 
deposition decreased exponentially with time which led to a gradual decrease in wind strength. 
In turn, as the wind abates, the outward motion of the reverse shock slows and, eventually, we find 
that its motion can be completely halted. This stalling of the reverse shock also occurred in the 
two-dimensional simulation with the shock stalling at different times for different polar angles. 
We also find that in both 1D and 2D simulations the density jump across the reverse shock 
decreases as the shock is about to turn around. 
After the reverse shock stalls both the 1D and 2D simulations indicate that 
the reverse shock feature then moves back towards the core. In figure (\ref{fig:rho.snv1d.8e35.t=1,1.5,2,2.5,3}) we 
see this stalling of the reverse shock and the backwards motion can be seen after comparing the profiles at 
$t=2.5\;{\rm s}$ and $t=3\;{\rm s}$. The reverse shock in this simulation 
actually reached it's furthest radial position at about $t=2.4\;{\rm s}$. 
Similar backwards motion for the reverse shock may be seen in figure (1) of 
Tom{\`a}s \emph{et al.} \cite{2004JCAP...09..015T}. 

Eventually the reverse 
shock reaches the core whereupon our simulations indicate that 
it is reflected and subsequently becomes a weak forward shock. 
\begin{figure}
\includegraphics[width=5in]
{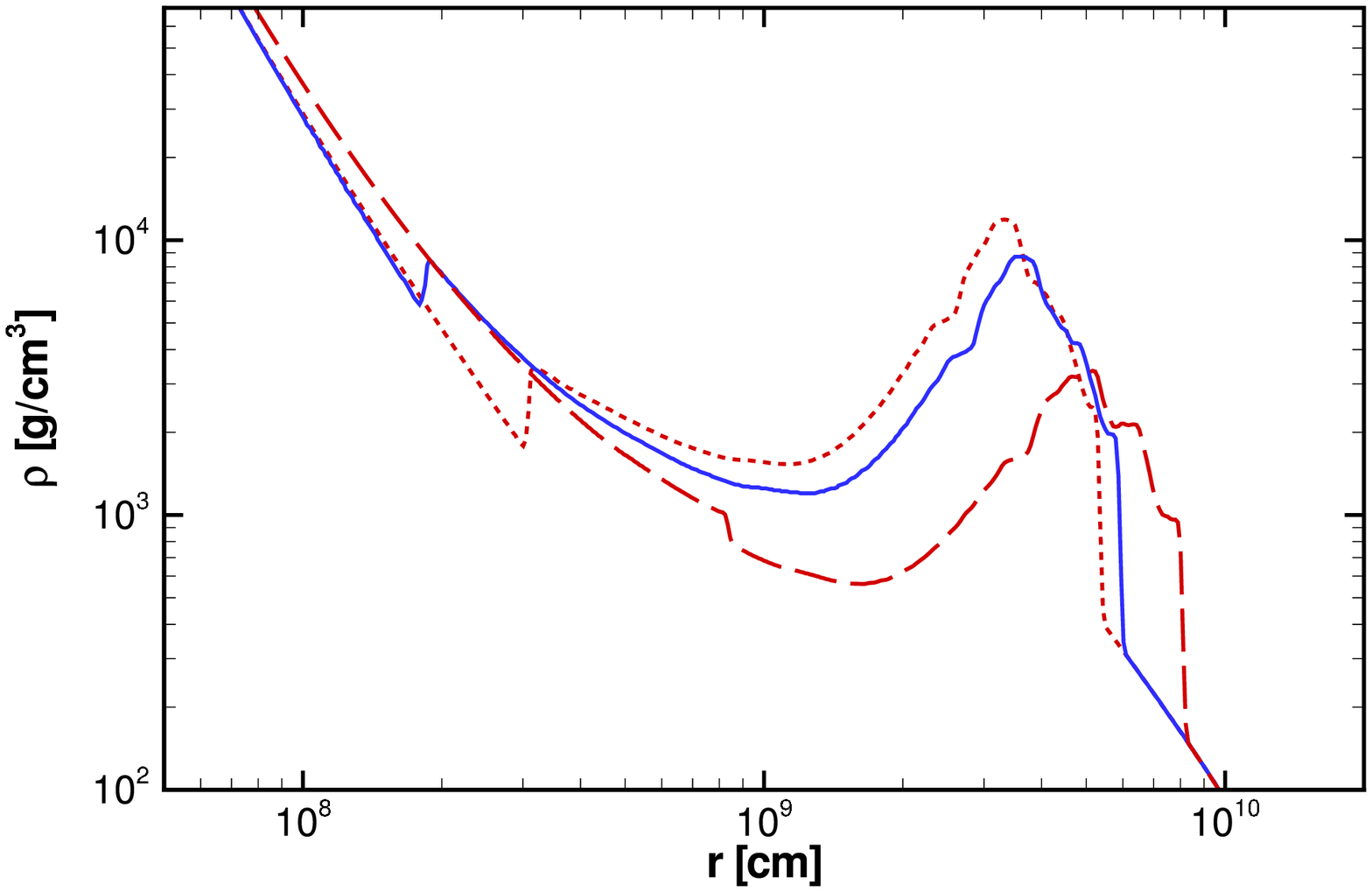}
\caption{\label{fig:rho.snv1d.8e35.t=2.8,3.1,4.3} The density as a function of the radius in the same 
1D SN model shown in figure (\ref{fig:rho.snv1d.8e35.t=1,1.5,2,2.5,3}) i.e. with 
$Q=3.36 \times 10^{51}\;{\rm erg}$, at $t=2.8\;{\rm s}$ (dotted), $t=3.3\;{\rm s}$ (solid) and $t=4.3\;{\rm s}$ (long dashed).}
\end{figure}
Further snapshots of the density profile 
taken from the simulation with $Q=3.36 \times 10^{51}\;{\rm erg}$ are shown in 
figure (\ref{fig:rho.snv1d.8e35.t=2.8,3.1,4.3}) where we see more clearly the backwards motion of the 
reverse shock and its later reflection. 
The radial position at which the reverse shock stalls depends upon the 
energy deposition. Additional snapshots from the simulation with slightly larger energy deposition, 
$Q=4.51 \times 10^{51}\;{\rm erg}$, are show in figure (\ref{fig:rho.snv1d.2e36.t=4,4.5,4.9}). 
Though the reverse shock in this simulation also attained it's maximum radial position at $t=2.4\;{\rm s}$ 
the figure shows that it was located further out before it was turned around. 
\begin{figure}
\includegraphics[width=5in]
{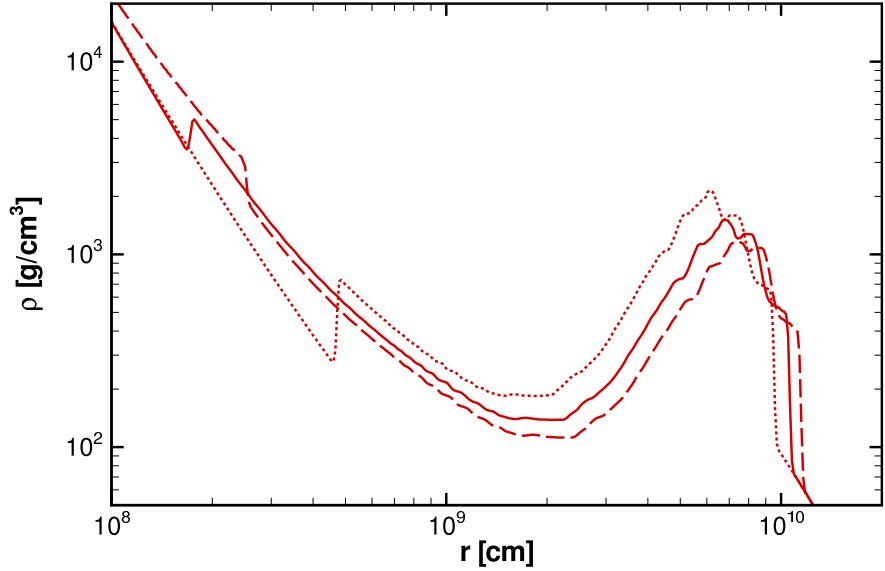}
\caption{\label{fig:rho.snv1d.2e36.t=4,4.5,4.9} The density as a function of the radius in the same 
1D SN model shown in figure (\ref{fig:rho.snv1d.2e36.t=1,1.5,2}), that is 
$Q=4.51 \times 10^{51}\;{\rm erg}$, at $t=4.0\;{\rm s}$ (dotted), $t=4.5\;{\rm s}$ (solid) and $t=4.9\;{\rm s}$ (long dashed).}
\end{figure}

The reverse shock is an interesting feature of the SN and, in contrast with the forward shock, can move both 
outwards and inwards. For very energetic explosion it may penetrate to lower densities than that immediately 
in front of the forward shock. The density jump across the reverse shock can vary with time and does so most 
noticeably when the shock is about to stall. 


\subsubsection{Asphericity}

The difference in the hydrodynamics between the one and two dimensional SN simulations 
can be traced back to the aspherical heating of the material above 
the proto-neutron star in the two dimensional case. Non-spherical heating can lead to turbulent fluid flow 
creating eddies and bubbles as shown in figure (\ref{fig:rho.snv2d.t=2.5}). 
\begin{figure}
\includegraphics[width=5in]
{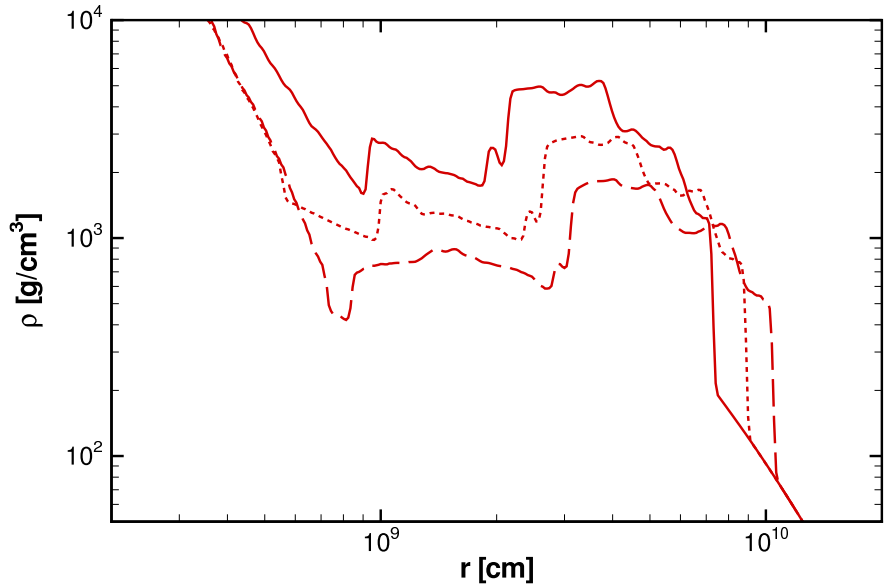}
\caption{\label{fig:rho.snv2d.t=3.9,4.8,5.7.slice=25} The density at a polar angle of $25^{\circ}$ 
as a function of the radius in a 2D SN model at $t=3.9\;{\rm s}$ (solid), $t=4.8\;{\rm s}$ (dotted) 
and $t=5.7\;{\rm s}$ (dashed).}
\end{figure}
\begin{figure}
\includegraphics[width=5in]
{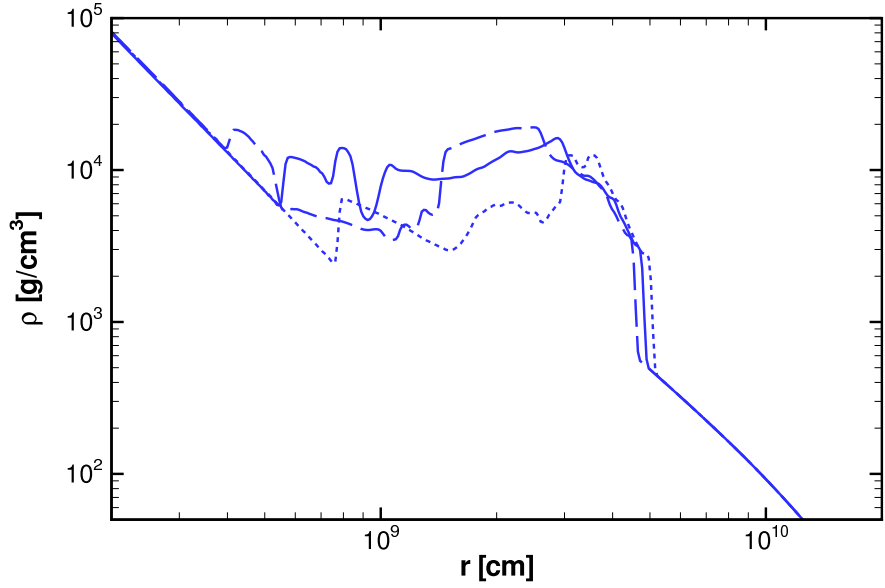}
\caption{\label{fig:rho.snv2d.t=2.5.slice=45,105,165} The density as a function of the radius in a 
2D SN model at $t=2.5\;{\rm s}$. The angular slices are at $45^{\circ}$ (solid), $105^{\circ}$ (dotted)
and $165^{\circ}$ (dash dot).}
\end{figure}
The neutrinos released by the proto-neutron star propagate along radial slices of the profile. In 
Fig. (\ref{fig:rho.snv2d.t=3.9,4.8,5.7.slice=25}) we show the density profile 
from the 2-D model at the polar angle of 
$\theta=25^{\circ}$ for snapshots at $t=3.9\;{\rm s}$, $t=4.8\;{\rm s}$ and $t=5.7\;{\rm s}$ while in 
figure (\ref{fig:rho.snv2d.t=2.5.slice=45,105,165}) we present the density along three different polar angles 
at $t=2.5\;{\rm s}$, i.e. taken from figure (\ref{fig:rho.snv2d.t=2.5}). 
In both we see the forward and reverse shocks and, as mentioned earlier, 
figure (\ref{fig:rho.snv2d.t=2.5.slice=45,105,165}) also shows that the 
radial position of both the forward and reverse shocks varies with the polar angle.
Figure (\ref{fig:rho.snv2d.t=3.9,4.8,5.7.slice=25}) also indicates that between 
$t=4.8\;{\rm s}$ and $t=5.7\;{\rm s}$ the reverse shock (along this radial slice) stalled and 
began to move back to the core. 
\begin{figure}
\includegraphics[width=5in]
{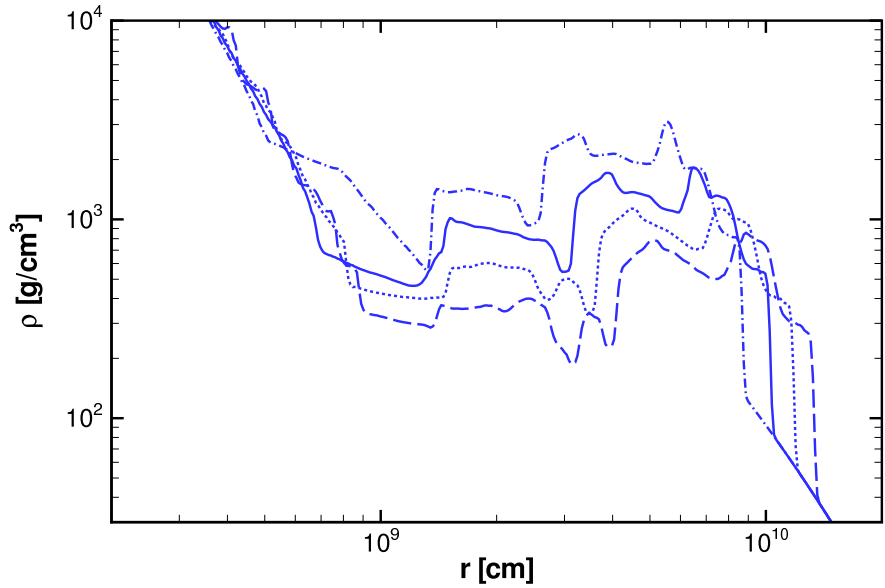}
\caption{\label{fig:rho.snv2d.t=4.5,5.4,6.3,7.2.slice=125} The density at a polar angle of $125^{\circ}$ 
as a function of the radius in a 2D SN model at $t=4.5\;{\rm s}$ (dash dot), $t=5.4\;{\rm s}$ (solid), 
$t=6.3\;{\rm s}$ (dotted), and $t=7.2\;{\rm s}$ (dashed).}
\end{figure}
For other radial slices the reverse shock turnaround time will be different: in figure 
(\ref{fig:rho.snv2d.t=4.5,5.4,6.3,7.2.slice=125}) we plot radial slices along the $125^{\circ}$ line of sight
at various times. The reverse shock in the figure, located just beyond $r = 10^{9}\;{\rm cm}$, turns around 
between $5.4\;{\rm s}$ and $6.3\;{\rm s}$. 

In all our 2D figures the general shape of the density profile between the two shocks is reminiscent 
of that shown in figure (\ref{fig:rho.snv1d.8e35.t=1,1.5,2,2.5,3}) for the 1D simulation with 
a similar value for $Q$ - i.e. a lower density region in front of the reverse shock, a wall of higher density 
material behind the forward shock - but clearly there are large `fluctuations' upon this basic trend both 
between the shocks and behind the reverse shock. 


\subsection{Summary}

The forward shock, the reverse shocks, contact discontinuities and the local bubbles/cavities 
are the features in the density profile with the greatest potential to alter the state of any neutrinos propagating 
through the SN since they represent the locations within the SN where the density gradient is 
largest. Large density gradients lead to non-adiabatic evolution of 
the neutrinos and significant differences compared to the neutrino propagation through 
the undisturbed profile. 


\section{Neutrino Mixing}

The vast majority of the neutrinos emitted by the supernova last interact with matter at the neutrinosphere 
located at the surface of the proto-neutron star. As they propagate outwards a small percentage will be absorbed by nucleons 
thereby transferring energy to the supernova and, it is thought, reviving the stalled shock. But even though 
the remainder of the neutrinos
survive the passage through the material overlying the proto-neutron star what emerges is not 
the same as what was emitted. The change in the neutrinos is due to neutrino oscillations and the 
presence of matter modulates this mixing. 

Neutrino oscillations arise due to a distinction between the interaction eigenstates (otherwise known as the 
flavor eigenstates $e, \mu, \tau$) and the eigenstates of the free Hamiltonian (known as the mass eigenstates) 
with masses $m_{1}, m_{2}, m_{3}$. Since there are three basis states a general neutrino wavefunction is 
described by three complex coefficients and evolves according to the Schrodinger equation. In the 
vacuum the Hamiltonian is diagonal in the mass basis but possesses off-diagonal terms in the flavor basis
that are the cause of flavor oscillations. In the presence of 
matter a potential, $V({\bf r})$, that 
takes into account coherent forward scattering of the neutrinos, is included in the Hamiltonian. Since 
we are only concerned with mixing between active neutrino flavors 
(i.e. all the flavors that have ordinary weak interactions) 
we may subtract off the common neutral current contribution (which will contribute only an overall phase)
leaving 
just the charged current contribution to the $\nu_{e}-\nu_{e}$ component of $V({\bf r})$. This contribution 
is the well-known $V_{ee}({\bf r})=\sqrt{2}\,G_F Y_{e}({\bf r})\,\rho({\bf r}) / m_{N}$ where $G_F$ is 
Fermi's constant, $\rho({\bf r})$ 
is the mass density, $Y_{e}({\bf r})$ is the electron fraction and $m_{N}$ is the nucleon mass. 
The effect of the matter upon the antineutrinos differs from that of 
the neutrinos - the potential $\bar{V}({\bf r})$ that 
appears in the antineutrino Hamiltonian has the same magnitude but the opposite sign.
Due to the inclusion of $V_{ee}({\bf r})$ the Hamiltonian is a function of position and
is neither diagonal in the mass basis nor the flavor basis. 
One may try and diagonalize $H({\bf r})$ but the unitary transformation that 
relates the flavor basis to the new `matter' basis also varies with 
the position. Consequently the gradient of the unitary transformation is non-zero and 
one finds that the Schrodinger equation in this new matter basis - that was meant to diagonalize $H$ - picks up 
off-diagonal terms. Thus it is not possible, in general, to find a basis that diagonalizes the 
Hamiltonian and therefore oscillations of the amplitudes that describe 
the wavefunction occur in every basis. 

In general, the three complex components of the wavefunction oscillate simultaneously. Exactly what occurs 
depends upon the energy of the neutrino $E$, the differences between the squares of the 
masses $m_{1}, m_{2}, m_{3}$, the mixing angles that describe the relationship between 
the flavor and mass basis, and, 
of course, $V_{ee}({\bf r})$. 
Since there are three mass eigenstates there are three separate mass splittings
$\delta m^{2}_{ij}=m_{i}^{2}-m_{j}^2$ (though only two are independent) and the 
relationship between the neutrino flavor and mass bases is described by the matrix $U$ which, in 
turn, is parameterized by three 
mixing angles $\theta_{12}, \theta_{13}$ and $\theta_{23}$ plus a CP-phase $\delta$. The structure of 
$U$ is 
\begin{equation}
U = \left(\begin{array}{lll} c_{12}\,c_{13} & s_{12}\,c_{13} & s_{13}\,e^{-\imath\delta} \\ -s_{12}\,c_{23}-c_{12}\,s_{13}\,s_{23}\,e^{\imath\delta} & c_{12}\,c_{23}-s_{12}\,s_{13}\,s_{23}\,e^{\imath\delta} & c_{13}\,s_{23} \\ s_{12}\,s_{23}-c_{12}\,s_{13}\,c_{23}\,e^{\imath\delta} & -c_{12}\,s_{23}-s_{12}\,s_{13}\,c_{23}\,e^{\imath\delta} & c_{13}\,c_{23} \end{array}\right)
\end{equation}
where $c_{ij} = \cos \theta_{ij}$, $s_{ij} = \sin \theta_{ij}$. 
Mixing has been observed in the neutrinos emitted by the Sun and the neutrinos 
produced by cosmic rays striking the atmosphere. Both have been confirmed with terrestrial experiments. 
Each observation of neutrino mixing can be described by a single $\delta m^{2}$ - $\theta$ pair of 
parameters and experimentally the `solar' and `atmospheric' mass splittings differ 
by around a factor of $\sim 30$ with the solar mass splitting, $\delta m_{\odot}$, being the smaller of the 
two. This observation 
permits us to consider the evolution of the general, three component, neutrino wavefunction as 
being factored into spatially distinct, two-neutrino mixes. The factorization simplifies matters greatly and 
from it one can demonstrate that there are two resonances in the supernova density profile: the 
so-called `L' resonance 
and the `H' resonance. The L resonance, at lower density and thus further from the proto-neutron star, 
involves mixing between matter 
states $\nu_1$ and $\nu_2$, the relevant mass splitting is $\delta m^{2}_{21}$ - which is 
approximately the solar mass splitting $\delta m_{\odot}$ - and the mixing angle 
is $\theta_{12}$ - which is approximately the 
mixing angle determined by the solar neutrino experiments $\theta_{\odot}$. The relevant 
mass eigenstates and mass splitting for 
the H resonance, at higher density and closer to the core, depend upon the sign of $\delta m^{2}_{32}$ (or, equivalently, $\delta m^{2}_{31}$) 
and this is not currently known. 
If $\delta m^{2}_{32}$ is positive (a normal hierarchy) then the H resonance involves 
mixing between states $\nu_2$ and $\nu_3$ and the mass splitting $\delta m^{2}_{32}$.  If $\delta m^{2}_{32}$ is 
negative (an inverted hierarchy) then at the H resonance it is the antineutrinos 
states $\bar{\nu}_1$ and $\bar{\nu}_3$ that mix and 
the relevant mass splitting is $\delta m^{2}_{31} = \delta m^{2}_{32} + \delta m^{2}_{21}$. 
In either case the mixing angle is the unknown, but small, $\theta_{13}$. 
Quite generally the coherent matter basis wavefunction that arrives  
at a distance, $d$, from the core of the SN
is related to the initial matter basis wavefunction at 
the proto-neutron star via the equation $\psi_{\nu}(d,E) = S_\nu (d,E) \,\psi_{\nu}(0,E)$ 
while for the anti-matter states $\psi_{\bar{\nu}}(d,E) = S_{\bar{\nu}} (d,E) \,\psi_{\bar{\nu}}(0,E)$.
The matrices $S_\nu$ and $S_{\bar{\nu}}$ are the S-matrices for the neutrinos and antineutrinos
respectively, for a discussion of this approach to 
neutrino oscillations, see e.g. \cite{Kneller:2005hf}.
To determine $\psi_{\nu}(d,E)$ and $\psi_{\bar{\nu}}(d,E)$ we need to know both the initial states $\psi_{\nu}(0,E)$
and $\psi_{\bar{\nu}}(0,E)$ and the two matrices $S_\nu$ and $S_{\bar{\nu}}$.

The matrices $S_\nu (d,E)$ and $S_{\bar{\nu}}(d,E)$ can be factored 
as $S_\nu(d,E)=S_{V,\nu} \, S_{L,\nu}(E) \, S_{H,\nu}(E)$ and
$S_{\bar{\nu}}(d,E)= S_{V,{\bar{\nu}}} \,S_{H,\bar{\nu}}(E)$ which are sufficiently general to accommodate our 
lack of knowledge of the hierarchy. 
The two matrices $S_{V,\nu}$ and $S_{V,\bar{\nu}}$ represent the 
neutrino propagation from the surface of the SN through the vacuum to Earth and both are diagonal in the mass basis. 
The matrices $S_{L,\nu}(E)$, $S_{H,\nu}(E)$ and $S_{H,{\bar{\nu}}}(E)$ represent the change to the initial wavefunction due 
to the neutrino's or antineutrino's passage through the H or L resonance\footnote{Many of the profiles presented 
in section \S\ref{section:simulations} have multiple H resonances therefore the matrices $S_{H,\nu}(E)$ and $S_{H,{\bar{\nu}}}(E)$ 
represent their combined effect of all the H resonances and $S_{L,\nu}(E)$ all the L resonances.  
We assume that the neutrinos
encounter all L-type resonances after all H-type resonances.  This is the case for all profiles we have generated}.
There is no matrix  
$S_{L,{\bar{\nu}}}(E)$ because antineutrinos do not experience an L resonance in matter.
The two matrices $S_{H,\nu}(E)$ and $S_{H,{\bar{\nu}}}(E)$ are evaluated somewhere between the H and L resonances 
while $S_{L}(E)$ is evaluated at the surface of the SN. These evaluation positions will not affect our 
result if they are sufficiently far from the resonances. 
The structures of $S_{L,\nu}(E)$, $S_{H,\nu}(E)$ and $S_{H,{\bar{\nu}}}(E)$ follow from our knowledge 
of the matter states that mix at either the H or L resonance and are thus
\begin{eqnarray}
S_{L,\nu}(E) & = & \left(\begin{array}{ccc} \alpha_{L}(E) & \beta_{L}(E) & 0 \\ -\beta_{L}^{\ast}(E) & \alpha_{L}^{\ast}(E) & 0 \\ 0 & 0 & 1 \end{array}\right) \\
S_{H,\nu}(E) & = & \left(\begin{array}{ccc} 1 & 0 & 0 \\ 0 & \alpha_{H}(E) & \beta_{H}(E) \\ 0 & -\beta_{H}^{\ast}(E) & \alpha_{H}^{\ast}(E)\end{array}\right) \label{eq:SHnu}\\
S_{H,{\bar{\nu}}}(E) & = & \left(\begin{array}{ccc} \bar{\alpha}_{H}(E) & 0 & \bar{\beta}_{H}(E) \\ 0 & 1 & 0 \\ -\bar{\beta}_{H}^{\ast}(E) & 0 & \bar{\alpha}_{H}^{\ast}(E)\end{array}\right)\label{eq:SHnubar} 
\end{eqnarray}
after omitting irrelevant phases. From these matrices we define 
\begin{eqnarray}
P_{L}(E) & = & 1- |\alpha_{L}(E)|^{2} = |\beta_{L}(E)|^{2}, \\
P_{H}(E) & = & 1- |\alpha_{H}(E)|^{2} = |\beta_{H}(E)|^{2}, \\
\bar{P}_{H}(E) & = & 1- |\bar{\alpha}_{H}(E)|^{2} = |\bar{\beta}_{H}(E)|^{2},
\end{eqnarray}
which are the crossing probabilities for neutrinos or antineutrinos at the two resonances. 
The resonances are said to be `adiabatic' or `non-adiabatic' depending upon whether the crossing probability 
is close to zero or closer to unity. Indeed these are the two natural values since the 
crossing probability is determined by the ratio of the resonance width to the local oscillation lengthscale
and typically one is significantly larger than the other.

Our interest now turns to the initial states. 
The density at the proto-neutron star is so large that the matter eigenstates and the 
flavor eigenstates are strongly aligned there. From a full 3-neutrino mixing calculation we find that 
the initial matter basis spectra for a normal hierarchy (NH) are 
$\Phi_{\nu_3}(0,E) = \Phi_{\nu_e}(0,E)$, $\Phi_{\bar{\nu}_1}(0,E) = \Phi_{\bar{\nu}_e}(0,E)$ while all other 
states, $\Phi_{{\nu}_1}(0,E), \Phi_{{\nu}_2}(0,E), \Phi_{\bar{\nu}_2}(0,E)$ and $\Phi_{\bar{\nu}_3}(0,E)$ are 
equal to the $\Phi_{\nu_\mu}(0,E),\Phi_{\nu_\tau}(0,E),\Phi_{\bar{\nu}_\mu}(0,E),\Phi_{\bar{\nu}_\tau}(0,E)$ spectrum which we 
call $\Phi_{\nu_x}(0,E)$. With an inverted hierarchy (IH) the initial states 
are $\Phi_{\nu_2}(0,E) = \Phi_{\nu_e}(0,E)$, $\Phi_{\bar{\nu}_3}(0,E) = \Phi_{\bar{\nu}_e}(0,E)$ and this time 
$\Phi_{\nu_1}(0,E),\Phi_{\nu_3}(0,E), \Phi_{\bar{\nu}_1}(0,E)$ and $\Phi_{\bar{\nu}_2}(0,E)$ are all 
equal to $\Phi_{\nu_x}(0,E)$.

Putting the initial spectra and definitions of the S-matrices together 
one finds that the flux of matter state $i$ a distance d from the supernova is given by
\begin{eqnarray}
F_{\nu_i}(d,E) & = & \frac{1}{4\pi\,d^{2}}\;\sum_{j} |{\left(S_{\nu}\right)}_{\, ij}(d,E)|^{2}\,\Phi_{\nu_j}(0,E), \label{eq:Fi} \\
F_{\bar{\nu}_i}(d,E) & = & \frac{1}{4\pi\,d^{2}}\;\sum_{j} |{\left(S_{\bar{\nu}}\right)}_{ij}(d,E)|^{2}\,\Phi_{\bar{\nu}_j}(0,E). \label{eq:Fbari}
\end{eqnarray}
where $\Phi_{\nu_i}(0,E)$ and $\Phi_{\bar{\nu}_i}(0,E)$ are the initial spectra of the 
matter states. 
But what are detected at Earth are, of course, the flavor states. During their flight from the 
supernova to Earth any coherence between the matter eigenstates is lost 
so the flux of flavor $\alpha$ that arrives at 
Earth is the incoherent sum 
\begin{equation}
F_{\nu_\alpha}(d,E) = \sum_{i} |U_{\alpha i}|^{2}\,F_{\nu_i}(d,E).
\end{equation}
These flavor fluxes may be rewritten in terms of the fluxes emitted by the neutrinosphere by introducing the survival 
probabilities $p(E)$ and $\bar{p}(E)$ for the electron neutrinos and antineutrinos respectively 
since these are the two flavors with distinct initial spectra. Written this way the detectable flavor fluxes are 
\begin{eqnarray} 
& & F_{\nu_e}(d,E) = \frac{1}{4\,\pi\,d^{2}}\;\left[\,p(E)\,\Phi_{\nu_{e}}(0,E) + (1-p(E))\,\Phi_{\nu_x}(0,E)\right], \\ 
& & F_{\bar{\nu}_e}(d,E) = \frac{1}{4\,\pi\,d^{2}}\;\left[\,\bar{p}(E)\,\Phi_{\bar{\nu}_e}(0,E) + (1-\bar{p}(E))\,\Phi_{\nu_x}(0,E)\right], \\
& & 4F_{\nu_x}(d,E) = \frac{1}{4\,\pi\,d^{2}}\;\left[ (1-p(E))\,\Phi_{\nu_e}(0,E) + (1-\bar{p}(E))\,\Phi_{\bar{\nu}_e}(0,E) + (2+p(E)+\bar{p}(E))\,\Phi_{\nu_x}(0,E)\right]\nonumber\\
& &
\end{eqnarray}
where we have adopted the notation of Dighe \& Smirnov \cite{Dighe:1999bi} by denoting $F_{\nu_\mu}+F_{\nu_\tau}+F_{\bar{\nu}_\mu}+F_{\bar{\nu}_\tau}$ by $4\,F_{\nu_x}$.
From equations (\ref{eq:Fi}) and (\ref{eq:Fbari}), and the initial spectra, we find that $p(E)$ and $\bar{p}(E)$
are related to the elements of $S(E)$ and $S_{\bar{\nu}}(E)$ via 
\begin{eqnarray}
p(E) & = & \sum_{i} |U_{ei}|^{2}\,|{\left( S_\nu \right)}_{ij}(E)|^{2} \\
\bar{p}(E) & = & \sum_{i} |U_{ei}|^{2}\,|{\left(S_{\bar{\nu}}\right)}_{ik}(E)|^{2}
\end{eqnarray}
where $j=3$, $k=1$ for a normal hierarchy, and $j=2$, $k=3$ for an inverted hierarchy. After inserting 
the definition of the crossing probabilities $p(E)$ and $\bar{p}(E)$ become 
\begin{eqnarray}
& & NH\;\; \left\{ \begin{array}{l} 
p(E) = |U_{e1}|^{2}\,P_{L}(E)\,P_{H}(E) + |U_{e2}|^{2}\,(1-P_{L}(E))\,P_{H}(E) + |U_{e3}|^{2}\,(1-P_{H}(E)) \\
\bar{p}(E) = |U_{e1}|^{2}\,(1-\bar{P}_{H}(E)) + |U_{e3}|^{2}\,\bar{P}_{H}(E) 
\end{array}\right.\nonumber \\
& & \label{eq:ppbarNH} \\
& & IH\;\;\left\{ \begin{array}{l} 
p(E) = |U_{e1}|^{2}\,P_{L}(E)\,(1-P_{H}(E)) + |U_{e2}|^{2}\,(1-P_{L}(E))\,(1-P_{H}(E)) + |U_{e3}|^{2}\,P_{H}(E) \\
\bar{p}(E) = |U_{e1}|^{2}\,\bar{P}_{H}(E) + |U_{e3}|^{2}\,(1-\bar{P}_{H}(E)) 
\end{array}\right. \nonumber \\
& & \label{eq:ppbarIH}
\end{eqnarray}
These formulae are quite general but from our knowledge of the various neutrino oscillation parameters 
we can be more specific about what exactly happens to the neutrinos and antineutrinos. 
The solar mixing parameters are such that the neutrino L resonance is almost completely 
adiabatic until, perhaps, the very end of the supernova neutrino signal. For this reason $P_{L}(E) = 0$. 
Thus at the end we find that determining the neutrino fluxes at Earth comes down to 
computing $P_{H}(E)$ or $\bar{P}_{H}(E)$. 

The Schrodinger equation forms a starting point 
by which $P_{H}(E)$ or $\bar{P}_{H}(E)$ can be determined. But if one na\"{i}vely applies, for example, a 
Runge-Kutta integrator to this equation one quickly discovers that this is a difficult problem 
from a numerical standpoint because such algorithms are simply not suited to differential 
equations where the solution is a highly oscillatory function. A number of alternate methods have 
been developed for calculating $P_{H}(E)$ or $\bar{P}_{H}(E)$, such as application of the the Landau-Zener result or 
the semi-analytic method by Balantekin \& Beacom \cite{BB1996}, but for one reason or 
another these alternate approaches can break down for complex profiles. 
Some authors \cite{Kneller:2005hf,ioa04,ioa05,akh05} have recognized that 
the evolution of the neutrino wave can be recast as a scattering problem and we adopt in this 
paper the algorithm outlined in Kneller \& McLaughlin \cite{Kneller:2005hf} which computes the evolution 
of the neutrino wavefunction by a Monte Carlo integration. For this paper we 
selected $|\delta m^{2}| = 3\times10^{-3}\,{\rm eV^{2}}$ and 
for $\theta_{13}$ we use $\sin^2\theta_{13} = 10^{-4}$ as a representative value. With this choice 
of $\theta_{13}$ the unperturbed profile is just sufficiently adiabatic to suppress the transformation 
effects of all but the most significant density features. The current experimental limit on $\theta_{13}$ is set
by the CHOOZ experiment \cite{REACTOR REVIEW}, and for our choice of $\delta m^{2}$ that limit
is $\sin^{2}\theta_{13} < 0.1$.

Finally, in addition to the effects caused by the propagation through the supernova there are potentially 
Earth matter effects that can arise. These are straightforward to compute and we do not consider them here since 
their presence (or absence) in the neutrino signal is a function of the position of the supernova with respect to the 
detector when the event occurs. 


\section{Profile features and the effects upon the crossing probability} \label{sec:pfcp}

With the SN simulations complete and a method for calculating the crossing probabilities in hand 
we turn to study the effects of various features in the profiles upon $P_{H}$ both as a function of 
time and energy. 


\subsection{The Forward Shock}

The forward shock is a generic feature of all our SN simulations. Initially the shock is located close to the
core and at high density and then, after it is regenerated, begins to move outwards and to lower densities as shown in 
the figures from section \S\ref{section:simulations}. 
\begin{figure}
\includegraphics[width=5in]
{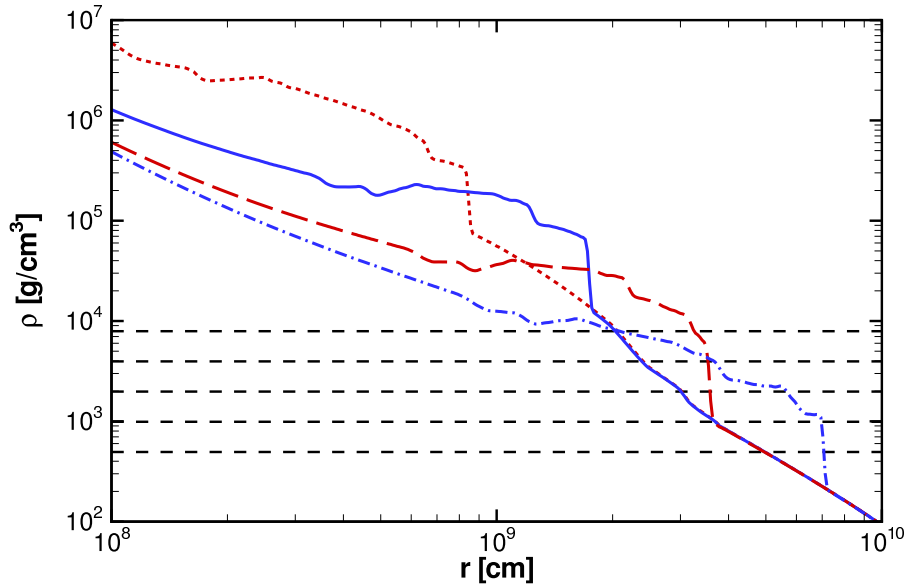}
\caption{\label{fig:rho.snv1d.1e35.t=0.9,1.8,3.6,7.2.+rhores} The density as a function of the radius in a 1D SN model 
with $Q=1.66 \times 10^{51}\;{\rm erg}$ at $t=0.9\;{\rm s}$ (dotted), $t=1.8\;{\rm s}$ (solid), 
$t=3.6\;{\rm s}$ (long dashed) and $t=7.2\;{\rm s}$ (dash dot). 
The horizontal dashed lines are (from top to bottom) the resonance densities for $5$, $10$, $20$, $40$ and 
$80\;{\rm MeV}$ neutrinos.}
\end{figure}
In Fig. (\ref{fig:rho.snv1d.1e35.t=0.9,1.8,3.6,7.2.+rhores}) we reproduce the results from 
Fig. (\ref{fig:rho.snv1d.1e35.t=0.9,1.8,3.6,7.2}) and also superimpose the resonance densities for 
$5$, $10$, $20$, $40$ and $80\;{\rm MeV}$ neutrinos. The radii where the horizontal dashed lines intercept the profile 
are the locations of the resonances. 
\begin{figure}
\includegraphics[width=5in]
{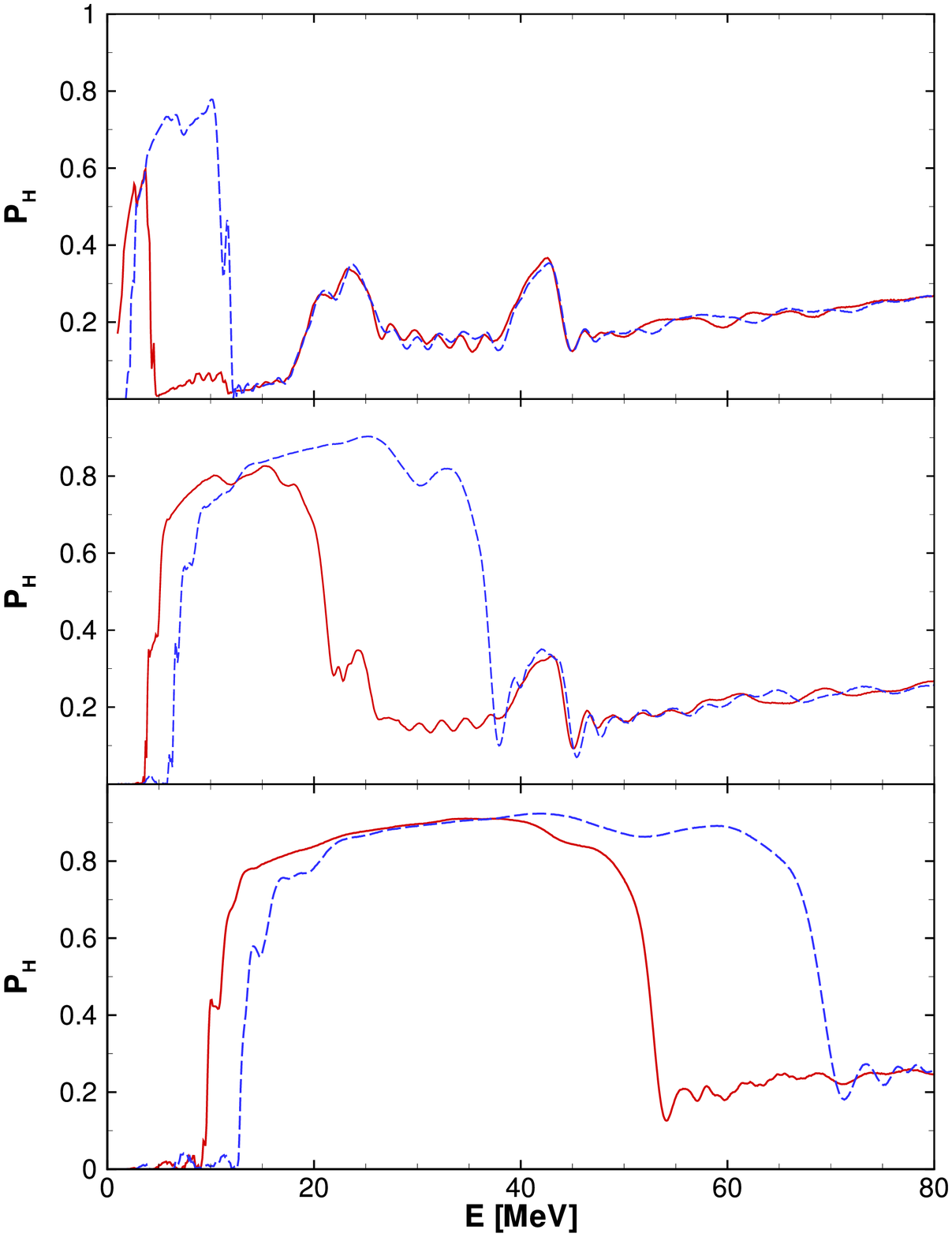}
\caption{\label{fig:PC.snv1d.1e35.t=2,2.5,3,3.5,4,4.5} The H resonance crossing 
probability $P_{H}$ as a function of neutrino energy 
for the 1D SN simulation with $Q=1.66 \times 10^{51}\;{\rm erg}$. In the top panel 
the snapshot times are $t=2\;{\rm s}$ (solid) and $t=2.5\;{\rm s}$ (dashed), in the middle panel $t=3\;{\rm s}$ (solid) 
and $t=3.5\;{\rm s}$ (dashed) and in the bottom panel $t=4\;{\rm s}$ (solid) and $t=4.5\;{\rm s}$ (dashed).}
\end{figure}
And in Fig. (\ref{fig:PC.snv1d.1e35.t=2,2.5,3,3.5,4,4.5}) we show $P_{H}$ as a function of neutrino energy at 
six snapshots of this same simulation. The crossing probability as a function of the energy clearly evolves with time. 
If $\theta_{13}$ is not too small then the evolution of the neutrinos through the undisturbed 
progenitor profile is almost adiabatic i.e. $P_{H} \sim 0$. As the 
forward shock arrives at the H-resonance for $5\;{\rm MeV}$ neutrinos - 
Fig. (\ref{fig:rho.snv1d.1e35.t=0.9,1.8,3.6,7.2.+rhores}) indicates this occurs at shortly after $t \sim 1.8\;{\rm s}$ - 
the evolution becomes non-adiabatic i.e. $P_{H} \sim 1$ because the density 
jump across the shock straddles the resonance densities of this neutrino energy. 
As time progresses and the shock moves outwards to lower densities we see from 
Fig. (\ref{fig:rho.snv1d.1e35.t=0.9,1.8,3.6,7.2.+rhores}) that the shock will begin to 
affect $40\;{\rm MeV}$ neutrinos at $t=3.6\;{\rm s}$.
The non-adiabaticity sweeps up through the neutrino spectrum from low energy to high. 
As time progress further eventually the shock ceases to affect the neutrinos of a particular energy 
and so their propagation returns to being adiabatic. For this simulation Fig. 
(\ref{fig:rho.snv1d.1e35.t=0.9,1.8,3.6,7.2}) indicates that the evolution of $5\;{\rm MeV}$ neutrinos returns
to being adiabatic at around $t=3.6\;{\rm s}$ and that for $40\;{\rm MeV}$ the return occurs roughly 
at $t=7.2\;{\rm s}$. Note that the $5\;{\rm MeV}$ neutrinos were affected for a much 
briefer period ($\sim 1.8\;{\rm s}$) than 
the $40\;{\rm MeV}$ neutrinos ($\sim 3.6\;{\rm s}$). This is due to the shape of the density profile. 
\begin{figure}
\includegraphics[width=5in]
{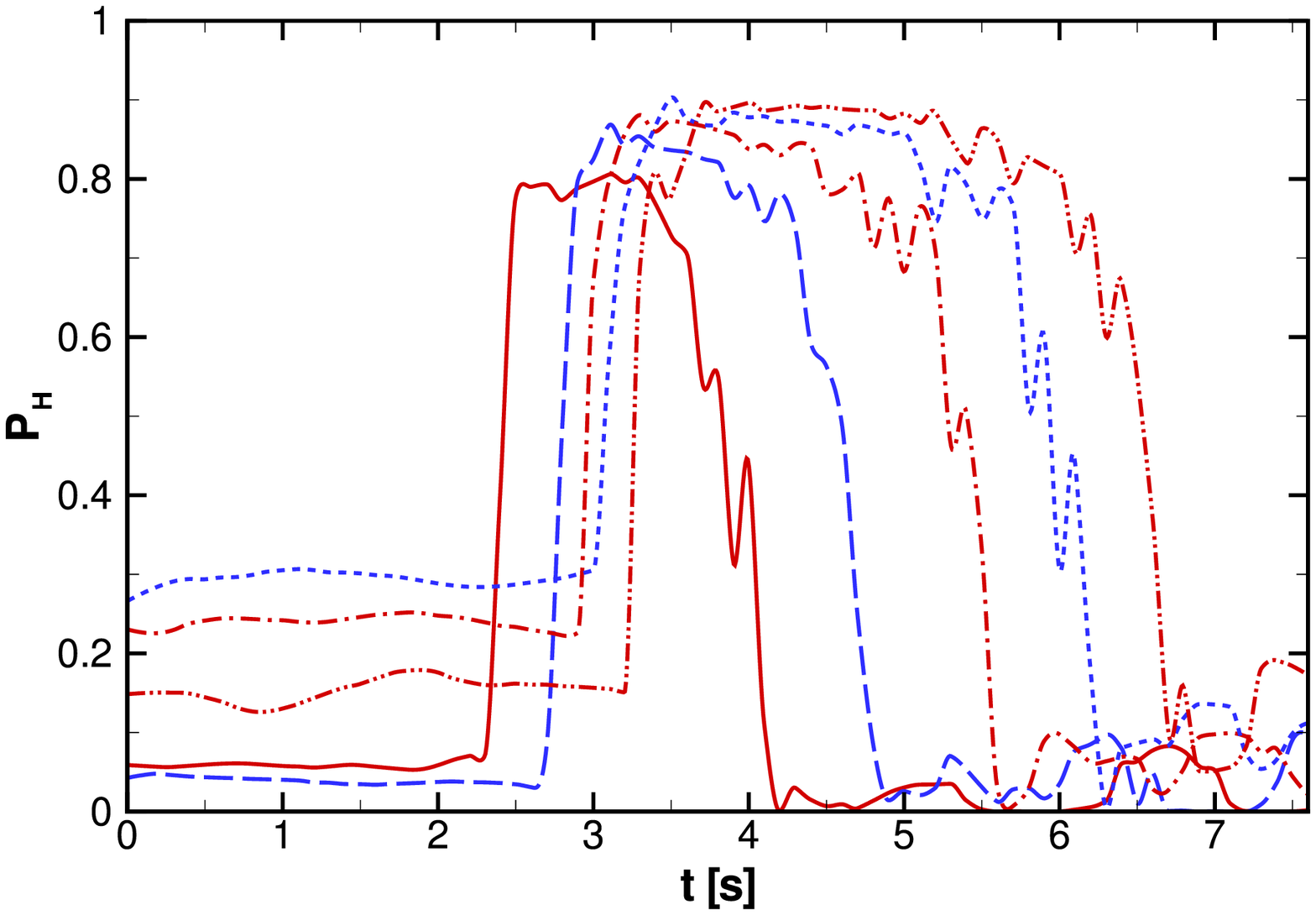}
\caption{\label{fig:PC.snv1d.1e35.E=10,15,20,25,30} The H resonance crossing 
probability $P_{H}$ as a function of time for selected neutrino energies through the 
1D SN model with $Q=1.66 \times 10^{51}\;{\rm erg}$. The curves are:
$E=10\;{\rm MeV}$ (solid), $E=15\;{\rm MeV}$ (long dashed), $E=20\;{\rm MeV}$ (dash-dot), $E=25\;{\rm MeV}$ (short dashed),
and $E=30\;{\rm MeV}$ (dash double-dot).}
\end{figure}
A complimentary perspective is to look at how particular neutrino energies evolve with time such as 
those in Fig. (\ref{fig:PC.snv1d.1e35.E=10,15,20,25,30}). 
The temporary transition to non-adiabaticity for each neutrino energy is clearly visible in the 
figure. This figure makes it most obvious that the lower energies are affected before 
the higher and also that the duration of the non-adiabatic period increases 
with the neutrino energy. 

The extent of the shock feature in neutrino energy seen in 
Fig. (\ref{fig:PC.snv1d.1e35.t=2,2.5,3,3.5,4,4.5}) is related to the density jump, 
$\Delta \rho$, across the shock. At any given time, if $E_{S}$ is the highest neutrino energy 
affected by the shock and $\Delta E$ is the range of neutrino energies 
then the density jump across the shock is 
\begin{equation}
\frac{\Delta \rho}{\rho} = \frac{\Delta E}{E_{S} - \Delta E}
\end{equation}
which is independent of the mixing parameters. From equation (\ref{eq:delta rho}) we saw that, for normal shocks, 
$\Delta \rho$ is a function of the Mach number and ratio of specific heats $\gamma$. If $M$ is large 
then $\Delta \rho$, and consequently $\Delta E$, are essentially just a function of $\gamma$ but 
if the Mach number $M$ is not too large - which may be the case - then the width of this feature could be used 
to infer $M$ if $\gamma$ is known. 
\begin{figure}
\includegraphics[width=5in]{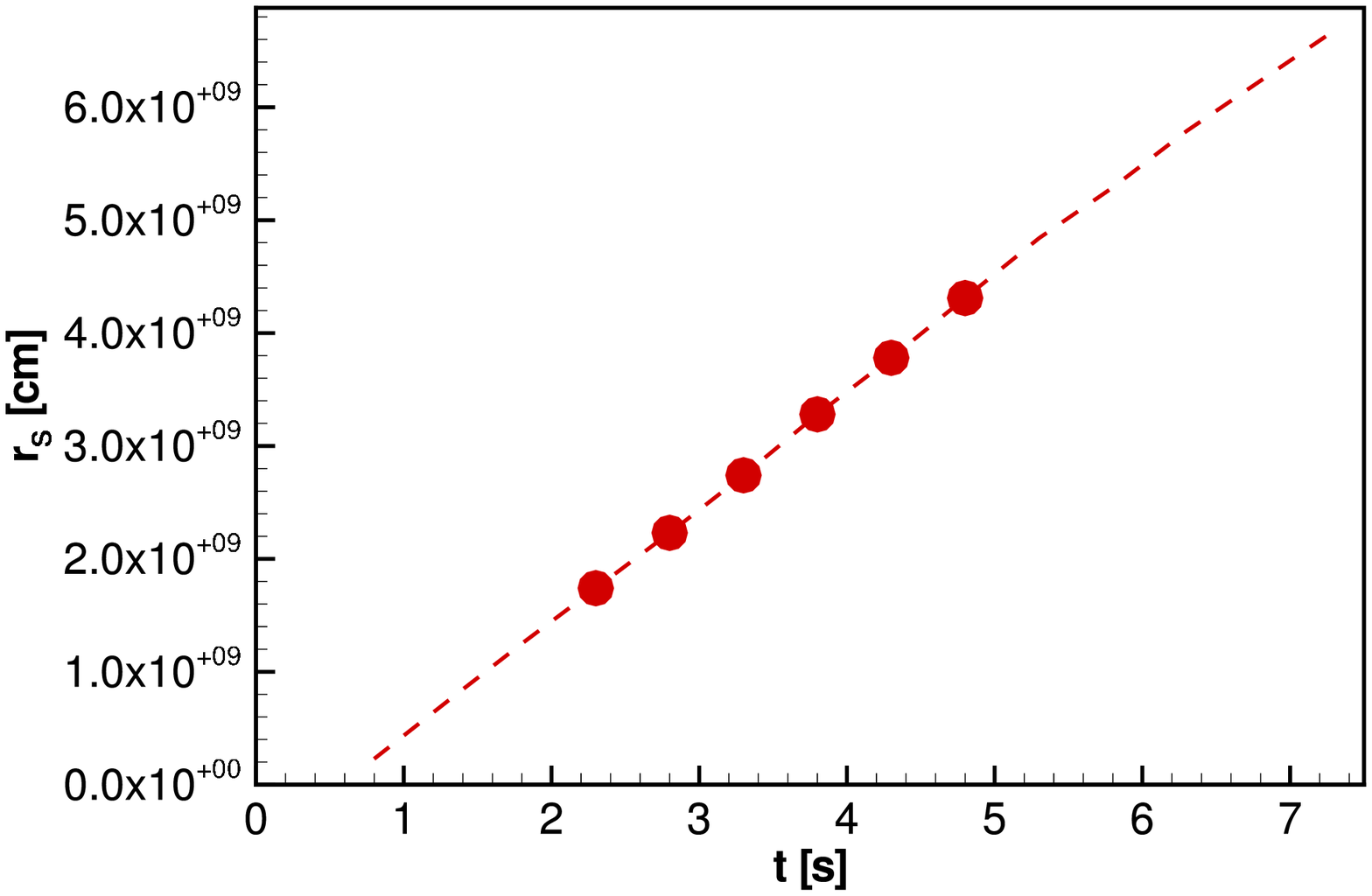} 
\caption{\label{fig:rs.snv1d.1e35} The shock position as a function of time, $r_{s}(t)$, 
for the 1D SN simulation where $Q=1.66 \times 10^{51}\;{\rm erg}$. An artificial 
time delay of $500\;{\rm ms}$ has been added to mimic the stalling of the shock at $r_{S} \sim 200\;{\rm km}$.}
\end{figure}
As the SN proceeds to explode $E_{S}$ moves 
up through the neutrino spectrum. The correspondence between energy and resonance density means that if we 
know, or assume, a progenitor profile shape then $E_{S}$ measures the shock position $r_{S}$. 
From the locations of $E_{S}$ shown in Fig. (\ref{fig:PC.snv1d.1e35.t=2,2.5,3,3.5,4,4.5}), 
and the initial profile, we show in Fig. (\ref{fig:rs.snv1d.1e35}) $r_{S}$ as function of time for 
this particular simulation.
The change of $r_{S}$ with time of course measures the shock speed but we 
can also use $r_{S}(t)$ to extrapolate the shock position both forward in time, if necessary, but, 
more interestingly, also backwards toward the proto-neutron star. If this extrapolation can be accomplished 
successfully then we can infer the moment, $t_{200}$, when $r_{S} \sim 200\;{\rm km}$. 
The neutronization burst (if detected) supplies a zero of time so $t_{200} \neq 0$ is 
the brief period the shock was stalled. 
Thus the neutrino signal, and in particular the location of the forward shock, contains 
evidence of whether a key component of the core-collapse 
supernova paradigm is correct and can provide quantitative data with which to compare with 
more sophisticated SN simulations than ours. 
We illustrate this idea in Fig. (\ref{fig:rs.snv1d.1e35}) where we have artificially included 
an offset in time. 


\subsection{The Reverse Shock}

The reverse shock was a feature seen in the more powerful explosions from section \S\ref{section:simulations} 
and the 2D result. This feature was not present in the initial profile and appeared later on when  
when the velocity of the wind superseded the local sound speed. After its formation the reverse shock initially 
moved outwards into the star but, as the energy deposition faded with time and the strength of the wind abated, 
eventually the reverse shock stalled and then headed back to the core. Like the 
forward shock, the reverse shock affects the adiabaticity of the evolution of the neutrino wavefunction 
through the profile so that neutrinos with resonance densities straddled by the 
density jump across the reverse shock evolve non-adiabatically as they pass through it. 
Note that any neutrino with a resonance density straddled 
by the density jump across the reverse shock will also experience two other resonances: one before the reverse shock 
and one after. 
\begin{figure}
\includegraphics[width=5in]
{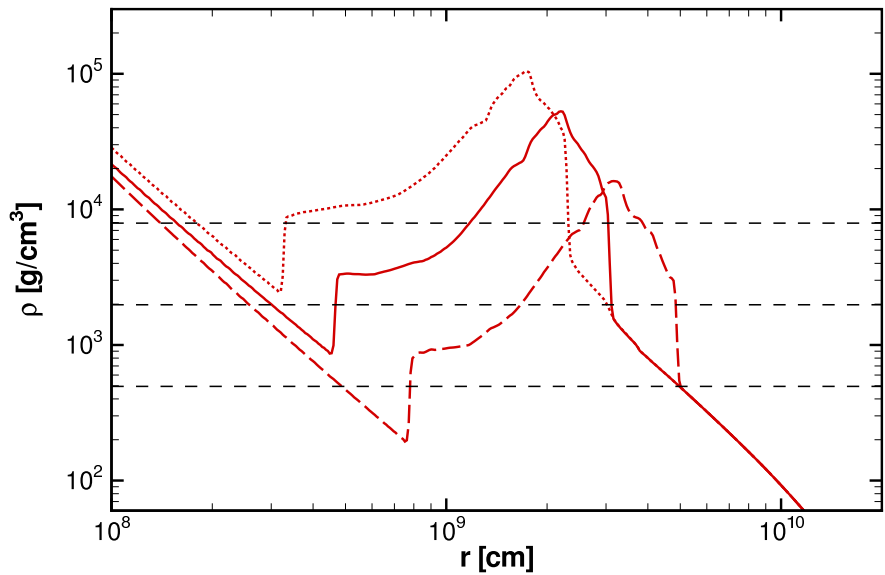}
\caption{\label{fig:rho.snv1d.2e36.t=1,1.5,2.+rhores} The density as a function of the radius in a 1D SN model
with $Q=4.51 \times 10^{51}\;{\rm erg}$ at $t=1.0\;{\rm s}$ (dashed), $t=1.5\;{\rm s}$ (solid) 
and $t=2.0\;{\rm s}$ (long dashed). The horizontal dashed lines are (from top to bottom) 
the resonance densities for $5$, $20$ and 
$80\;{\rm MeV}$ neutrinos.}
\end{figure}
This is shown in Fig. (\ref{fig:rho.snv1d.2e36.t=1,1.5,2.+rhores}) where we have superimposed 
the resonance densities for $5$, $20$, and $80\;{\rm MeV}$ neutrinos upon the profiles shown 
in Fig. (\ref{fig:rho.snv1d.2e36.t=1,1.5,2}). We see from the figure that it is possible for 
some neutrinos to be affected by both shocks. 
If this occurs, and if we neglect the effects from other resonances in the profile, 
then we might expect that the net effect upon such doubly shock-affected neutrinos to be zero \cite{2004JCAP...09..015T}.
\begin{figure}
\includegraphics[width=5in]
{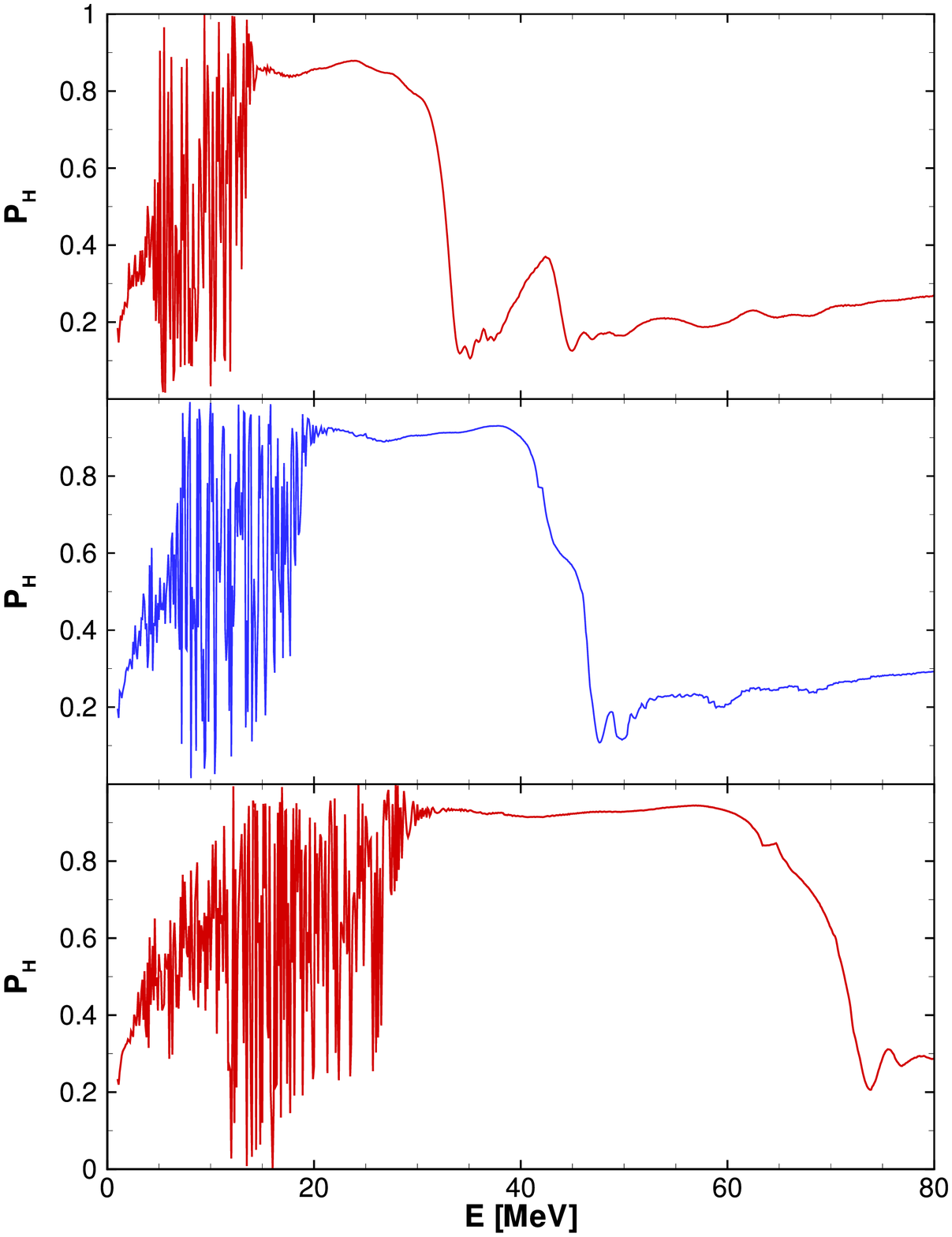}
\caption{\label{fig:PC.snv1d.8e35.t=1.8,2,2.4} The H resonance crossing probability 
$P_{H}$ as a function of neutrino energy for three snapshots taken from the 
1D SN model where $Q=3.36 \times 10^{51}\;{\rm erg}$. In the top panel the time is 
$t=1.8\;{\rm s}$, in the middle $t=2.0\;{\rm s}$, and in the bottom panel $t=2.4\;{\rm s}$.}
\end{figure}
However this expected cancellation is not seen in our results for $P_{H}$ as 
a function of the neutrino energy shown in Fig. (\ref{fig:PC.snv1d.8e35.t=1.8,2,2.4}) for 
the model where $Q=3.36 \times 10^{51}\;{\rm erg}$. 
The crossing probabilities plotted in the figure show some similarities to 
those plotted in Fig. (\ref{fig:PC.snv1d.1e35.t=2,2.5,3,3.5,4,4.5}) at the higher neutrino 
energies affected by the forward shock where there is change from adiabatic to non-adiabatic evolution. 
But for those energies where cancellation is naively expected we see instead that $P_{H}$ oscillates wildly. 
These rapid oscillations are phase effects due to the interference between the two shocks. 
Similar rapid oscillations in the crossing probability have been seen previously in SN neutrino calculations by 
Fogli \emph{et al.} \cite{Fetal2003} in a profile with a forward shock 
and then a bubble cavity behind it, in the SN test case used by Kneller \& McLaughlin \cite{Kneller:2005hf}, and in the 
results of Dasgupta \& Dighe \cite{Dasgupta:2005wn} where their significance was emphasized.

\begin{figure}
\includegraphics[width=5in]
{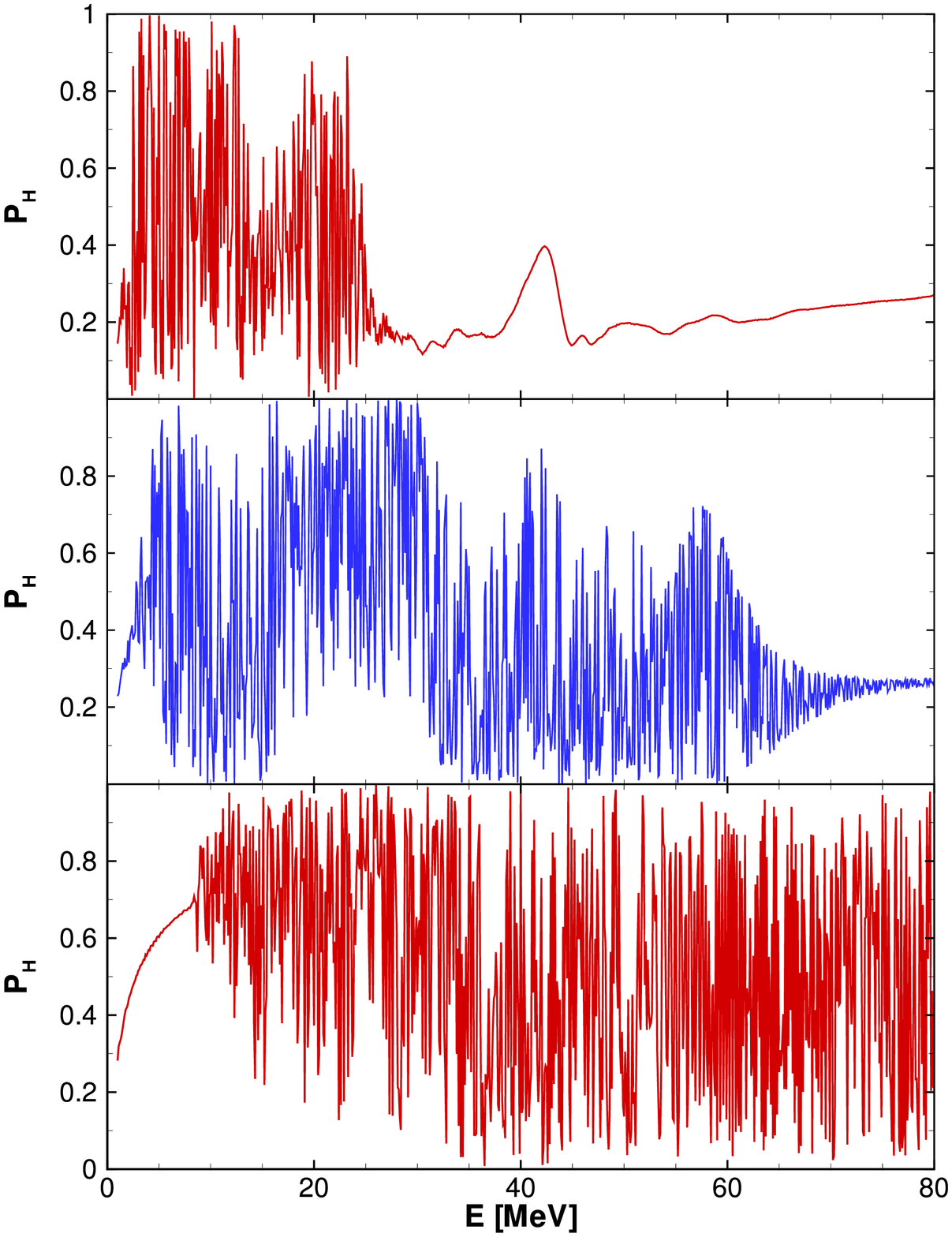}
\caption{\label{fig:PC.snv1d.2e36.t=1.1,1.4,3.0} The H resonance crossing probability 
$P_{H}$ as a function of neutrino energy for three snapshots taken from the 
1D SN model where $Q=4.51 \times 10^{51}\;{\rm erg}$. In the top panel the time is 
$t=1.1\;{\rm s}$, in the middle $t=1.4\;{\rm s}$, and in the bottom panel $t=3.0\;{\rm s}$.}
\end{figure}
Fig. (\ref{fig:PC.snv1d.2e36.t=1.1,1.4,3.0}) we display the crossing probability for 
the model where $Q=4.51 \times 10^{51}\;{\rm erg}$ at 
$t=1.1\;{\rm s}$, $t=1.4\;{\rm s}$ and $t=3.0\;{\rm s}$. Again phase effects 
are seen. Compared to the results shown in Fig. (\ref{fig:PC.snv1d.8e35.t=1.8,2,2.4}) 
for the weaker explosion where $Q=3.36 \times 10^{51}\;{\rm erg}$ and 
in Fig. (\ref{fig:rs.snv1d.1e35}) for 
$Q=1.66 \times 10^{51}\;{\rm erg}$, in this case there is no 
indication of the characteristic transition from adiabatic to non-adiabatic propagation associated with 
the forward shock. This occurs because, as noted earlier about this model, the reverse shock 
penetrates to lower densities than the forward shock at these early times. For this particular model 
the reverse shock affects the adiabaticity of a particular neutrino energy before the forward shock. 

The presence of phase effects in $P_{H}$ are caused by interference between forward and 
reverse shocks. But it would be nice to 
find in the neutrino signal a clean signature of the reverse 
shock that is not contaminated by the forward shock or other features of the profile. 
A potential signature could arise from the behavior noted earlier which is that the 
reverse shock stalled as the energy deposition into the 
material above the proto-neutron star faded and then headed back towards the core. 
As the reverse shock moves backwards its effects move down through the neutrino spectrum and might 
eventually cease to overlap with the forward shock. 
Since the density jump across the reverse shock becomes quite small by the time this behavior occurs 
the reverse shock will produce a narrow spectral feature. 
For the simulation where $Q=3.36 \times 10^{51}\;{\rm erg}$, 
shown in Fig. (\ref{fig:rho.snv1d.8e35.t=1,1.5,2,2.5,3}),
this expected pattern is difficult to see in the neutrino signal because the forward shock has not 
swept through the neutrino spectrum by the time the reverse shock turns around. But in 
the simulation with slightly larger energy deposition, $Q=4.51 \times 10^{51}\;{\rm erg}$, 
the forward shock moves much more quickly and has largely 
swept through the H resonance region by the time the reverse shock starts to make its way back to the core. 
For this simulation the backwards moving reverse shock is more visible in the neutrino signal. 
The crossing probability for this simulation at the snapshots shown in 
Fig. (\ref{fig:rho.snv1d.2e36.t=4,4.5,4.9}) are presented in 
Fig. (\ref{fig:PC.snv1d.2e36.t=4,4.5,4.9}). The backwards moving reverse shock is responsible for the 
phase effects at $E_{\nu} \sim 10\;{\rm MeV}$ in the middle panel. 
And then once the reverse shock has reached the core our boundary conditions there led to its reflection and so the 
reverse shock became a outward moving, weak forward shock and re-appeared in the neutrino spectrum as a 
spectral feature moving from low to high neutrino energies. This is the cause of the 
bump in $P_{H}$ seen at $E_{\nu} \sim 15-20\;{\rm MeV}$ in the bottom panel. 
\begin{figure}
\includegraphics[width=5in]
{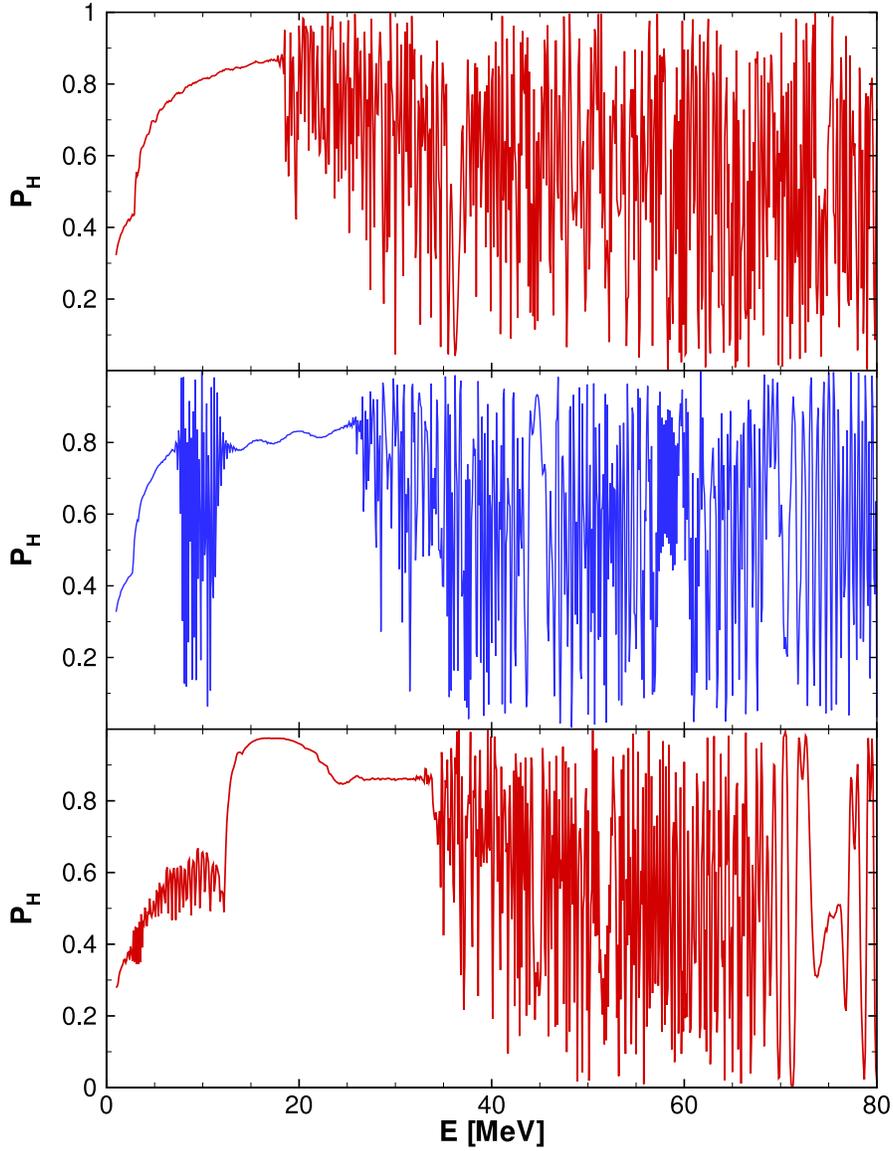}
\caption{\label{fig:PC.snv1d.2e36.t=4,4.5,4.9} The H resonance crossing probability $P_{H}$ as a 
function of neutrino energy for the 1D SN model where $Q=4.51 \times 10^{51}\;{\rm erg}$. 
From top to bottom the snapshot times are $t=4\;{\rm s}$, $t=4.5\;{\rm s}$ and $t=4.9\;{\rm s}$.}
\end{figure}

In summary, though it is apparent that the neutrino signal can vary considerably depending upon 
exactly how the reverse shock behaves the presence of strong phase effects - rapid oscillations 
with large amplitude in both time and energy of the crossing probability - 
are a notable signature of the presence of multiple shocks. Therefore, there exists tremendous 
potential in a future detection of the a supernova neutrino signal for probing the inner hydrodynamics 
of the explosion.


\subsection{Asphericity}

The notable difference between the one-dimensional and the two-dimensional models was the local density 
fluctuations that appeared in the latter. These local density enhancements/cavities affect 
only neutrinos along particular lines of sight so if one could determine their presence in the 
signal we would obtain an indication of the asphericity of the explosion. The most obvious method would be to 
compare different lines of sight to the same SN but obviously this is not practical. Instead 
we must find another signature of asphericity that one might use. 
\begin{figure}
\includegraphics[width=5in]
{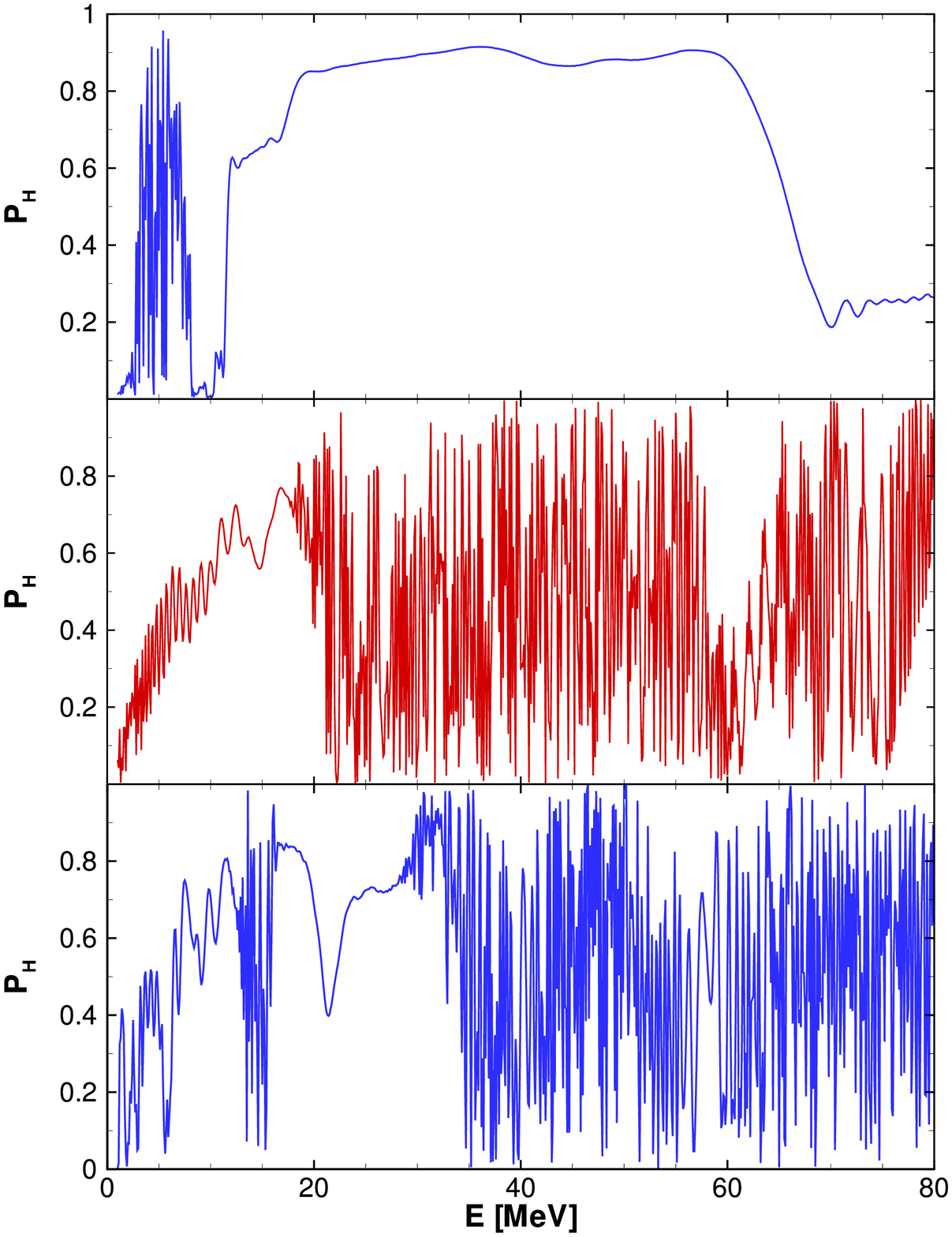}
\caption{\label{fig:PC.snv2d.t=2.4,5.4,6.4} The H resonance crossing probability $P_{H}$ as a 
function of neutrino energy for a radial slice at $\theta=25^{\circ}$ through the 2D SN model. 
From top to bottom the snapshot times are $t=2.4\;{\rm s}$, $t=5.4\;{\rm s}$ and $t=6.4\;{\rm s}$.}
\end{figure}
In figure (\ref{fig:PC.snv2d.t=2.4,5.4,6.4}) we show the crossing probability 
$P_{H}$ as a function of neutrino energy for the 
snapshot $t=2.4\;{\rm s}$, $t=5.4\;{\rm s}$ and $t=6.4\;{\rm s}$ and a line of sight $\theta=25^{\circ}$. 
Features in the figure resemble those seen in the 1D results: the forward shock leads to a range 
of energies with $P_{H}\rightarrow 1$ as seen in the top panel, phase effects due to presence of the 
reverse shock lead to the high `frequency' oscillations of $P_{H}$ seen in the 
middle panel, and the turn around of the reverse shock can be seen in the bottom panel at $E \sim 15\;{\rm MeV}$ 
where it appears, as before, as a narrow range of high frequency phase effects moving down through the spectrum.
While a detailed analysis of these results might indicate a difference that is due to asphericity 
there is no striking feature that one can point to.  
\begin{figure}
\includegraphics[width=5in]
{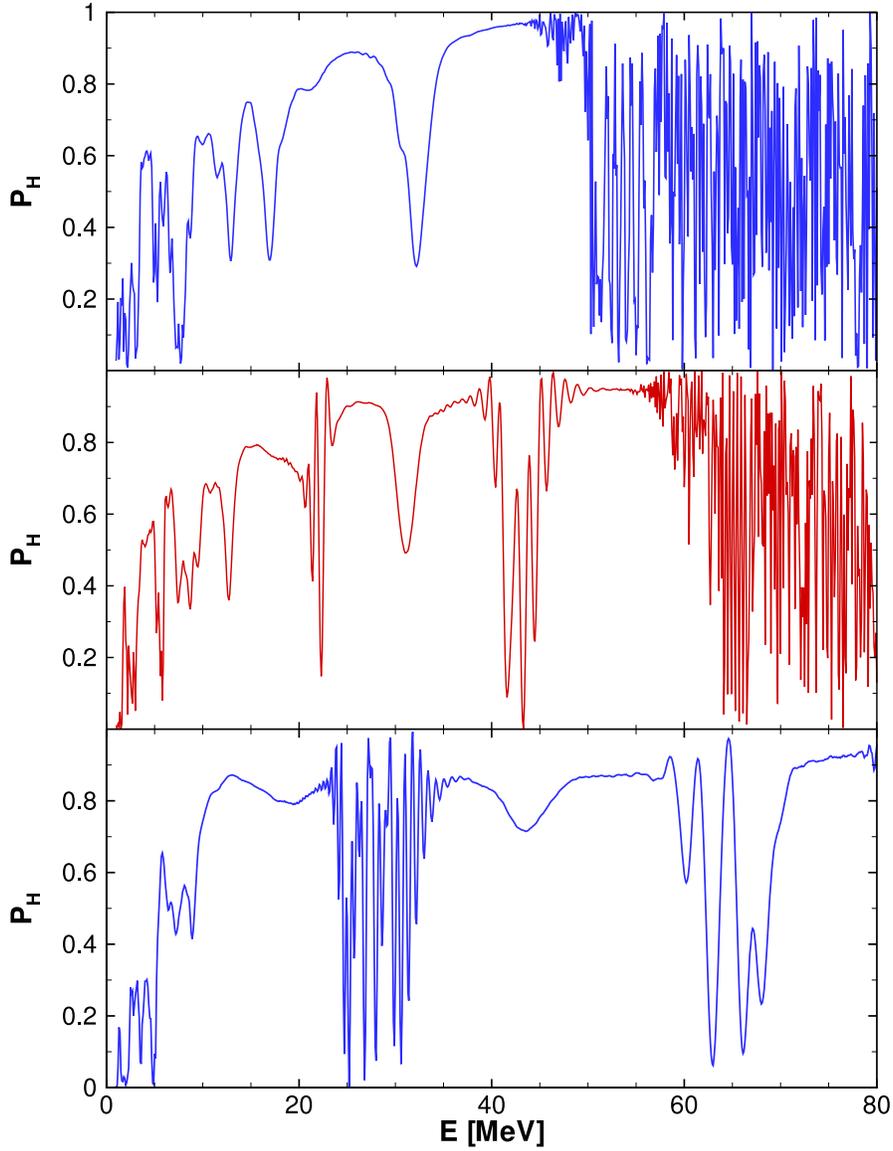}
\caption{\label{fig:PC.snv2d.t=7.4,8,9} The H resonance crossing probability $P_{H}$ as a 
function of neutrino energy for a radial slice at $\theta=25^{\circ}$ through the 2D SN model. 
From top to bottom the snapshot times are $t=7.4\;{\rm s}$, $t=8.0\;{\rm s}$ and $t=9.0\;{\rm s}$.}
\end{figure}
However, at late times we 
found a change in the qualitative appearance of $P_{H}$ versus 
neutrino energy. 
The high frequency phase effects no longer dominate the entire spectrum and now lower frequency 
portions show up. We did not see anything comparable at late times in any of our 1D simulation results. 
The changing quality of the phase effects are seen in Fig. (\ref{fig:PC.snv2d.t=7.4,8,9}) 
which shows the crossing probability at the H resonance as a function of neutrino energy for the same 
line of sight at $\theta=25^{\circ}$ through the 2D model. High frequencies can be seen above $E\sim 45\;{\rm MeV}$ 
in the top panel and above $E\sim 55\;{\rm MeV}$ in the middle panel but they have disappeared 
by $t=9.0\;{\rm s}$. In all three panels we see that at lower neutrino energies 
the curve is smoother with only narrow patches of high frequency phase effects that moved up through the spectrum. 
The shocks have largely ceased affecting the lower neutrino energies by this 
time so it is the `fluctuations' upon the profile - both between the shocks and behind the reverse shock -
which produce mildly adiabatic resonances, and/or resonances that are located very close to one another 
(and possibly even overlap) that give rise to these features in the figure. 

The effect upon neutrino flavor transport of supernovae density fluctuations 
was considered by Loreti et al \cite{Loreti:1995ae} and later by 
Fogli \emph{et al} \cite{2006JCAP...06..012F} and Friedland \& Gruzinov \cite{2006astro.ph..7244F}. 
These authors suggest that $P_{H}$ should tend to 1/2 in the presence of many small scale fluctuations. 
We note, however, that due to the exponential growth of the radial grid spacing the scale of features 
we can resolve by $r \sim 10^{4}\;{\rm km} - 10^{5}\;{\rm km}$ is of order $\delta r \sim 100\;{\rm km}$ 
and this is somewhat larger than the scale considered by Fogli \emph{et al}  
and Friedland \& Gruzinov. 
In our formulation of neutrino mixing using S-matrices we can motivate this expected result of 
$P_{H} \rightarrow 1/2$ by dividing the density profile into domains such 
that within each there is a single neutrino resonance. 
The passage of a neutrino (or antineutrino) through each interval is described 
by an S-matrix which has exactly the 
same structure as equations (\ref{eq:SHnu}) or (\ref{eq:SHnubar}) as appropriate. 
The magnitudes of the elements $\alpha$ and $\beta$ lie between 
zero and unity depending upon the adiabaticity of the resonance in that region.
The multiplication of all these S matrices together (in the correct order) to achieve the net 
effect of passing through all the resonances will tend to produce a net 
result where either $\alpha_{H}$ and $\beta_{H}$ or 
$\bar{\alpha}_{H}$ and $\bar{\beta}_{H}$ are of equal magnitude. 
In our simulations we had cases where we found up to 13 resonances 
but have not been able to reproduce the anticipated limiting behavior. 
Nevertheless, as shown in 
Figs. (\ref{fig:PC.snv2d.t=2.4,5.4,6.4}) and (\ref{fig:PC.snv2d.t=7.4,8,9}), 
we see a clear distinctive qualitative change in the
crossing probabilities at late time in aspherical models.


\section{Detector Signals}

Since the detection of neutrinos from SN 1987A in IMB \cite{1987PhRvL..58.1494B} and 
Kamiokande \cite{1987PhRvL..58.1490H} the number and size of the neutrino detectors have both 
increased in scale so that 
many thousands or even tens of thousands of neutrinos will be detected from the next Galactic supernova. 
Though many neutrino detectors are water Cerenkov detectors that are best suited to detecting the 
electron antineutrinos from supernovae via the inverse-$\beta$ reaction on protons, 
other neutrino detection technologies have 
been demonstrated or explored that allow one to see the electron neutrinos and the $\mu$/$\tau$ flavors 
via nucleus-neutrino reactions such as $O(\nu_{e},e)F$, or neutral current reactions such as 
$D(\nu_{X},\nu_{X})np$. The detection of these other neutrino flavors would provide a great deal of 
valuable information we could then use to learn about the SN explosion. 
For the purposes of demonstrating the temporal evolution of supernovae neutrino signals 
we shall consider two detector types: the 
water Cerenkov detector because the majority of 
neutrino detectors are of this type, and a heavy water detector (such as was used in the
Sudbury Neutrino Observatory)
since this detector was also considered by Schirato \& Fuller \cite{SF2002}. 
The largest neutrino detectors are water Cerenkov detectors. These detectors are sensitive to both 
electron neutrinos and electron anti neutrinos via the reactions $p(\bar{\nu}_{e},\bar{e})n$ and 
$O(\nu_{e},e)F$ \cite{1987PhRvD..36.2283H} with the inverse-$\beta$ reaction dominating the signal. 
The cross sections for the neutrino-nucleon reactions are well known but a good deal of uncertainty 
exists in the important neutrino-nucleus cross sections.
A heavy water detector is sensitive to both charged-current and neutral current events so 
could be used to detect the $\mu$ and $\tau$ flavors. However obtaining 
energy resolution for the neutral current events is difficult so the observable quantity 
considered by Schirato \& Fuller \cite{SF2002} was the ratio of total charged-current to 
neutral current events. Though temporal evolution of this ratio is somewhat smothered due to the fact 
that the event rates are integrated over the neutrino spectrum Schirato \& Fuller demonstrated 
that variations of the ratio would occur at the level of $10-20\%$. 
Lead detectors are also capable of detecting the $\mu$/$\tau$ flux and 
have been studied by Fuller, Haxton \& McLaughlin \cite{FHM1999}
Engel, McLaughlin \& Volpe \cite{Engel:2002hg} and Kolbe et al \cite{Kolbe:2000np}.

To determine what any given detector will see we must 
fold in three effects: the initial neutrino spectra emitted by the proto-neutron 
star, the energy dependence of the cross sections in the detector and, thirdly, 
the energy resolution of the detector. 


\subsection{Initial Spectra}

There is a considerable degree of uncertainty about the emitted neutrino spectra. 
Furthermore, neutrino self interactions - described in Samuel \cite{S1993}, 
Pastor, Raffelt \& Semikoz \cite{PRS2002} and Friedland \& Lunardini \cite{Friedland:2003dv} -
in the region immediately above the proto-neutron star
can alter the spectra after it's emission from the proto-neutron star but before the flux reaches 
the H resonance layer. There is a possibility that in some cases 
the spectra may be `equilibrated' by the time 
they get to the H resonance \cite{FQ2006,Sawyer:2005jk}. In such circumstances 
the changes in the density profile and the neutrino mixing 
described in previous sections would not lead to 
any observable changes in a detector except for the gradual fading of the signal.
We shall consider the case where such equilibration does not occur and 
experiment with 
an `alpha fit' power law for the spectrum exiting the neutrinosphere 
motivated both by its analytic simplicity and by recent supernova core 
Monte Carlo simulations \cite{2003ApJ...590..971K}. For the time dependence of the luminosity 
we use an exponential decay with a time constant of $\tau=3\;{\rm s}$.
The emitted differential spectra are then 
\begin{equation}
\Phi_{\nu}(0,E) = \frac{L_{\nu}\,(\alpha+1)^{(\alpha+1)}}{\langle E_{\nu}\rangle^{2}\,\Gamma(\alpha+1)}
\left(\frac{E}{\langle E_{\nu}\rangle}\right)^{\alpha}\,\exp\left[-(\alpha+1)\,\frac{E_{\nu}}{\langle E_{\nu}\rangle}\right]\,\exp[-t/\tau]. \label{eq:powlaw}
\end{equation}
where $L_{\nu}$ is the luminosity, $\langle E_{\nu}\rangle$ the mean energy and 
$\alpha$ is the `pinch' parameter. We adopt a hierarchy of mean energies: 
$\langle E_{\nu_{e}} \rangle =12\;{\rm MeV}$, $\langle E_{\bar{\nu}_{e}} \rangle =15\;{\rm MeV}$ 
and $\langle E_{\nu_{x}} \rangle =18\;{\rm MeV}$
(where, again, $x$ stands for either $\mu, \tau, \bar{\mu}$ or $\bar{\tau}$) and the luminosities 
are $L_{\nu_{e}}=L_{\bar{\nu}_{e}}=6\times 10^{52}\;{\rm erg/s}$, $L_{\nu_{x}}=2\times 10^{52}\;{\rm erg/s}$. 
\begin{figure}
\includegraphics[width=5in]
{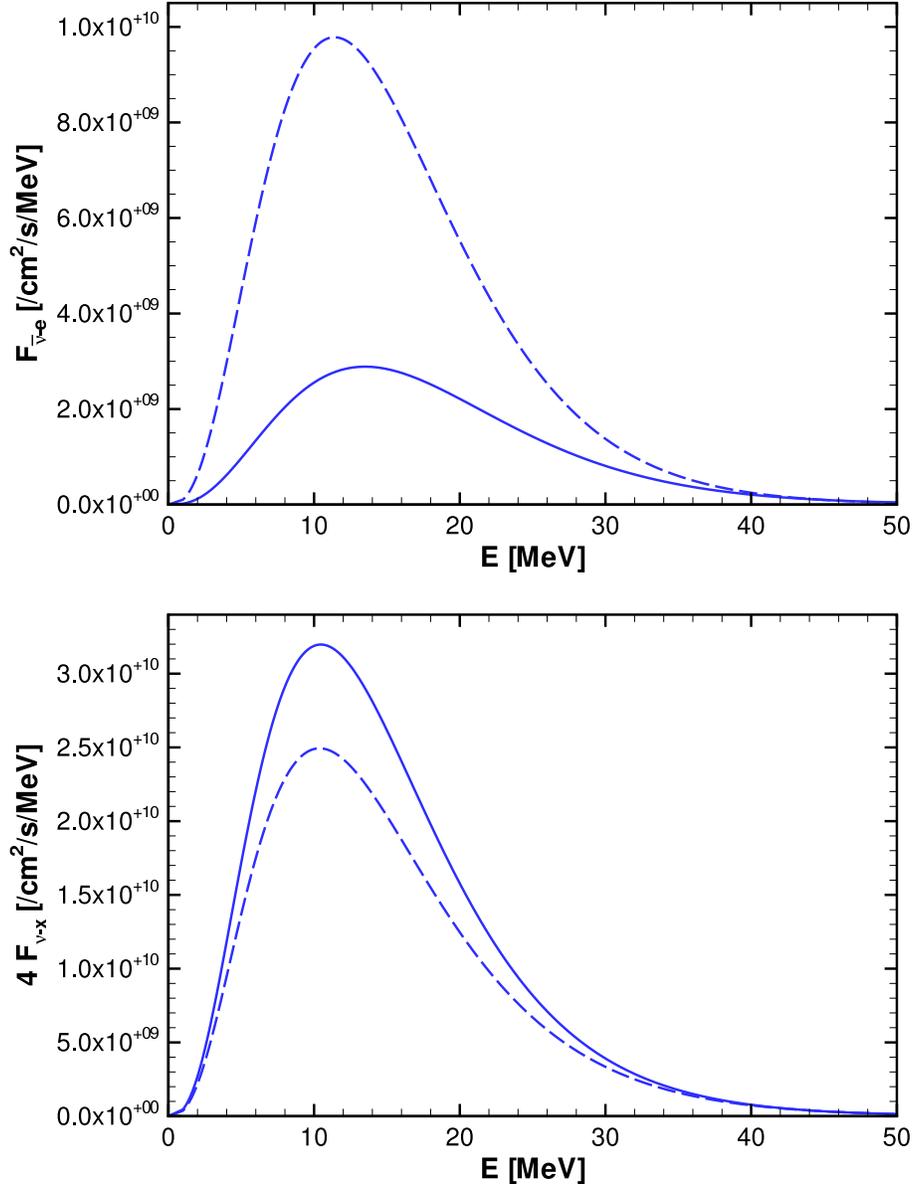} 
\caption{\label{fig:Fnu.alpha=3.IH.PHbar=0,1} 
Fluxes of electron antineutrinos (top panel) and the sum 
of $\nu_{\mu}+\bar{\nu}_{\mu}+\nu_{\tau}+\bar{\nu}_{\tau}$ neutrino fluxes (bottom panel) for 
completely adiabatic, i.e. $P_H =0$ (solid) and non-adiabatic, i.e. $P_H = 1$ (dashed) evolution. The hierarchy is inverted, 
the pinch parameter was set to $\alpha=3$ and the vacuum mixing angles are 
$\sin^{2}(2\theta_{12})=0.8$, $\sin^{2}(2\theta_{13})=4\times10^{-4}$. The distance to the SN was set at 
$d=10\;{\rm kpc}$ and the time $t$ to $t=0.0\;{\rm s}$.}
\end{figure}
\begin{figure}
\includegraphics[width=5in]
{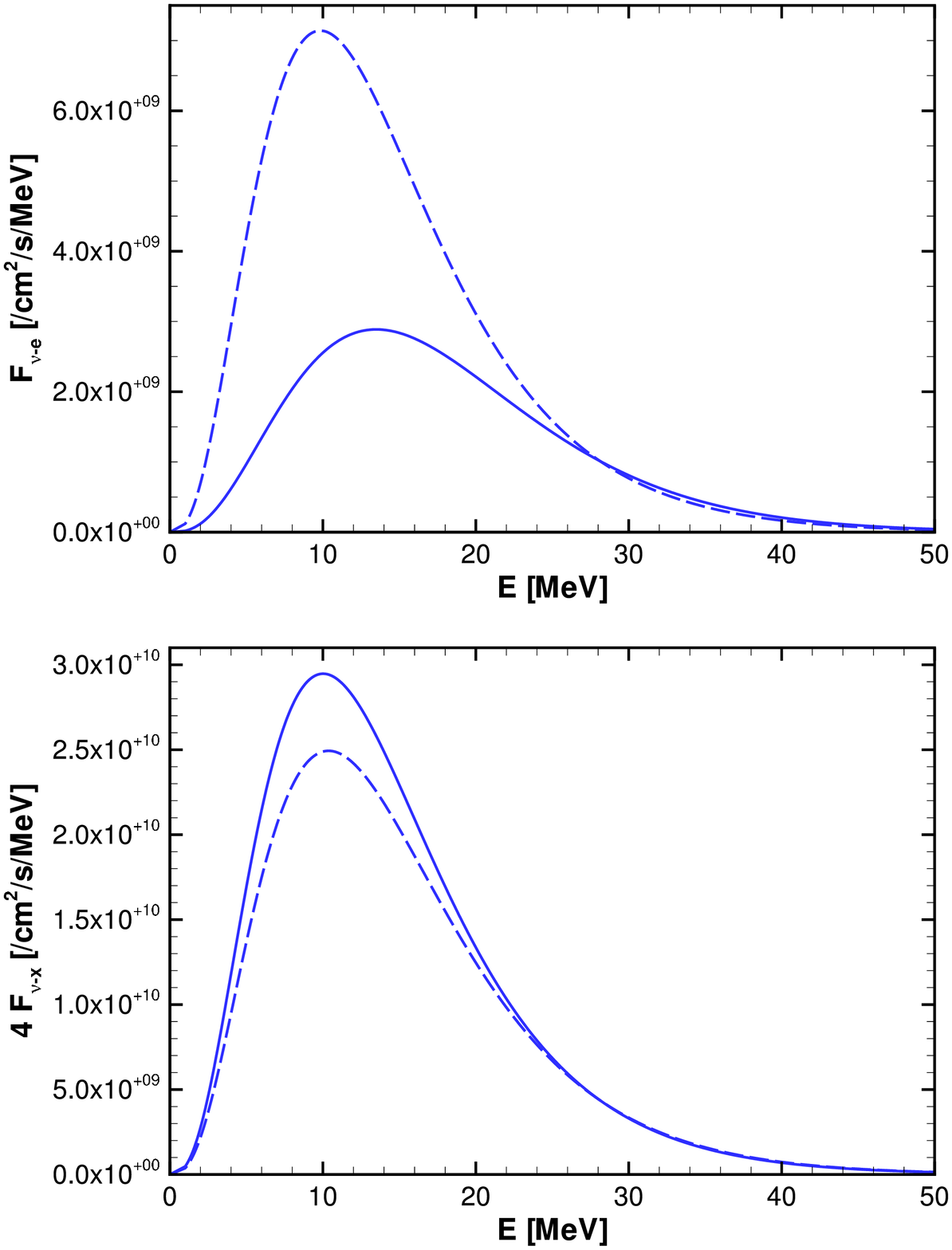}
\caption{\label{fig:Fnu.alpha=3.NH.PH=0,1} The fluxes of electron neutrinos (top panel) and the sum 
of $\nu_{\mu}+\bar{\nu}_{\mu}+\nu_{\tau}+\bar{\nu}_{\tau}$ neutrino fluxes (bottom panel) for 
completely adiabatic, i.e. $P_H =0$  (solid line) and non-adiabatic, i.e. $P_H = 1$ (dashed) 
evolution. Here the hierarchy is normal but 
otherwise all parameters are the same as those used in Fig. (\ref{fig:Fnu.alpha=3.IH.PHbar=0,1}).}
\end{figure}
In Figs. (\ref{fig:Fnu.alpha=3.IH.PHbar=0,1}) and (\ref{fig:Fnu.alpha=3.NH.PH=0,1}) 
we present the neutrino spectra that would be observed at Earth from a SN at a distance of $10\;{\rm kpc}$ 
if the evolution were either completely adiabatic or completely non-adiabatic 
for example through a single H resonance, i.e. $P_H = 0$ or $P_H = 1$ in either the inverted or normal hierarchy. 
The two curves in each panel of the figures therefore represent extremes in the sense that the actual 
flux at a given energy would fall somewhere in between the two curves. We do 
not plot the electron antineutrino flux for 
the normal hierarchy or the electron neutrino flux for the inverted hierarchy because these fluxes 
are always adiabatic and would only fade exponentially with time. 
In each panel of the two figures the pinch parameter was set to $\alpha=3$, other than a change 
in shape 
the general behavior of the curves 
at other values of $\alpha$ is unchanged. 
There are many points to notice in each figure. 
The first is 
that for any given neutrino flux and hierarchy there is a particular energy where the 
two curves cross. This will occur as long as the average energies of $\nu_x$ flavor neutrinos
are greater than that of the $\nu_{e}$ and $\bar{\nu}_e$. 
The crossing can be seen best in the $\bar{\nu}_{e}$ spectrum in the normal hierarchy. 
We define the energies where the curves cross as the critical energies. The critical energies 
vary with both flavor and the hierachy. 
At the critical energy for a particular hiearachy and flavor the adiabaticity 
of the neutrinos as they propagate through the profile is irrelevant and 
the flux of that flavor and at that energy through a detector is independent of $P_{H}$.
The other significant feature to notice is the large difference between the two curves for 
the electron neutrinos and electron antineutrinos for energies below $E \sim 30\;{\rm MeV}$.
This wide separation of the extremal fluxes at low to medium neutrino energies, $E \lesssim 30\;{\rm MeV}$, suggests 
that we focus our attention upon 
this portion of the spectrum because it is here that changes in adiabaticity will 
lead to the most significant changes in the flux. The flux at higher energies will also 
change with the evolving adiabaticity but to a lesser degree because the fluxes are so small.
Thus if the evolution of the neutrinos evolves from being adiabatic 
to non-adiabatic then, in the case of a normal hierarchy, the flux of electron neutrinos 
will increase for those energies below the critical energy and decrease above.  Similar 
behavior will occur for antineutrinos in the inverted hierarchy. The 
decrement above the critical energy is difficult to see in the figures because 
the critical energies are 
all far into the tail of the spectra. For the 
$\nu_\mu + \bar{\nu}_{\mu} + \nu_\tau + \bar{\nu}_\tau$ 
neutrino fluxes the opposite change occurs: a transition to non-adiabaticity leads to lower fluxes of 
these flavors at energies below the critical energy and higher fluxes above. 


\subsection{General Considerations}

Having determined the effects of the evolving density profile upon the 
neutrino mixing we are now in a position to examine the actual neutrino signal 
within a detector.
As discussed in section \ref{sec:pfcp}, the actual neutrino flux of a particular 
flavor through a detector differs from that emitted from the neutrinosphere. In the simplest scenario, 
if the neutrinos propagate through the 
outer layers of the SN adiabatically then the survival probabilities $p$ and $\bar{p}$ 
are found by setting $P_{H}=P_{L} = \bar{P}_{H}=0$ in equations (\ref{eq:ppbarNH}) and 
(\ref{eq:ppbarIH}). 
For the normal hierarchy $p=|U_{e3}|^{2}$ and $\bar{p}=|U_{e1}|^{2}$: for the inverse 
hierarchy $p=|U_{e2}|^{2}$ 
and $\bar{p}=|U_{e3}|^{2}$. Using the value of $\sin^{2} 2\theta_{13} = 4\times 10^{-4}$ 
we used to calculate $P_{H}$ and $\sin^{2} 2\theta_{12} = 0.8$ based upon the solar mixing parameters, 
we see that for the normal hierarchy the amount of the emitted $\nu_e$ flux present in the 
electron neutrino flux at a detector here on Earth is minuscule. Similarly we see that 
for an inverted hierarchy the amount of the emitted electron antineutrino flux present 
in any detected electron antineutrino flux is equally small. 
Thus if the mass hierarchy is normal, $\sin^{2} 2\theta_{13}$ is not too small, and 
the neutrino propagation adiabatic then the electron neutrino detectors on Earth will
essentially be detecting the emitted $\nu_\mu$ and $\nu_\tau$ flux. Or if the hierarchy is inverted, 
again $\sin^{2} 2\theta_{13}$ is not too small, and 
the neutrino propagation adiabatic then the electron antineutrino flux at Earth is the emitted 
$\bar{\nu}_\mu$ and $\bar{\nu}_\tau$. 
In order to see the emitted electron neutrino or antineutrino flux
we would need to detect the $\nu_\mu$, $\bar{\nu}_\mu$, $\bar{\nu}_\tau$ and/or $\nu_\tau$ flavors 
or seek the small amount 
of the emitted $\nu_e$/$\bar{\nu}_e$ present in the detected $\nu_{e}$/$\bar{\nu}_{e}$ flux.
In either case, it will be very difficult to reconstruct the the $\nu_e$ or $\bar{\nu}_e$ spectrum
that was originally emitted from the proto-neutron star.

The evolution of the density profile changes this conclusion. Depending upon 
the scenario, either $p$ (for a 
normal hierarchy) or $\bar{p}$ (for an inverted hierarchy) will change with time and 
so the proportions of the emitted 
fluxes present in the fluxes at Earth will also change. Note that $p$ and $\bar{p}$ do not 
both evolve with time within in the same scenario; 
one will stay constant. 
This means there is a potential to separate the time evolution 
of the emitted flux from the neutrinosphere from the evolution of the density profile through which 
the neutrinos propagate.  
As the star explodes and the forward shock eventually reaches the
H resonances the passage of portions of the neutrino spectrum 
through the density profile becomes non-adiabatic. Whether it is the neutrinos or antineutrinos 
that are affected is determined by the hierarchy. When the passage of the neutrinos is completely
non-adiabatic we see that $p$ or $\bar{p}$ change to $p=|U_{e2}|^{2}$ or $\bar{p}=|U_{e1}|^{2}$. 
Again using $\sin^{2} 2\theta_{13} = 10^{-4}$ and $\sin^{2} 2\theta_{12} = 0.8$ we see that 
$p \sim 72\%$, $\bar{p} \sim 28 \%$.
In effect 
the forward shock of such explosions acts like a window sweeping 
across either the neutrino or antineutrino spectra through which we observe an 
altered mixture of the emitted components. 
For neutrino or antineutrino energies below the 
critical energies this leads to an enhanced $\nu_{e}$ or $\bar{\nu}_{e}$ flux as shown in Figs.  
(\ref{fig:Fnu.alpha=3.IH.PHbar=0,1}) and (\ref{fig:Fnu.alpha=3.NH.PH=0,1}).


\subsection{A Heavy Water Detector}
\begin{figure}
\includegraphics[width=5in]
{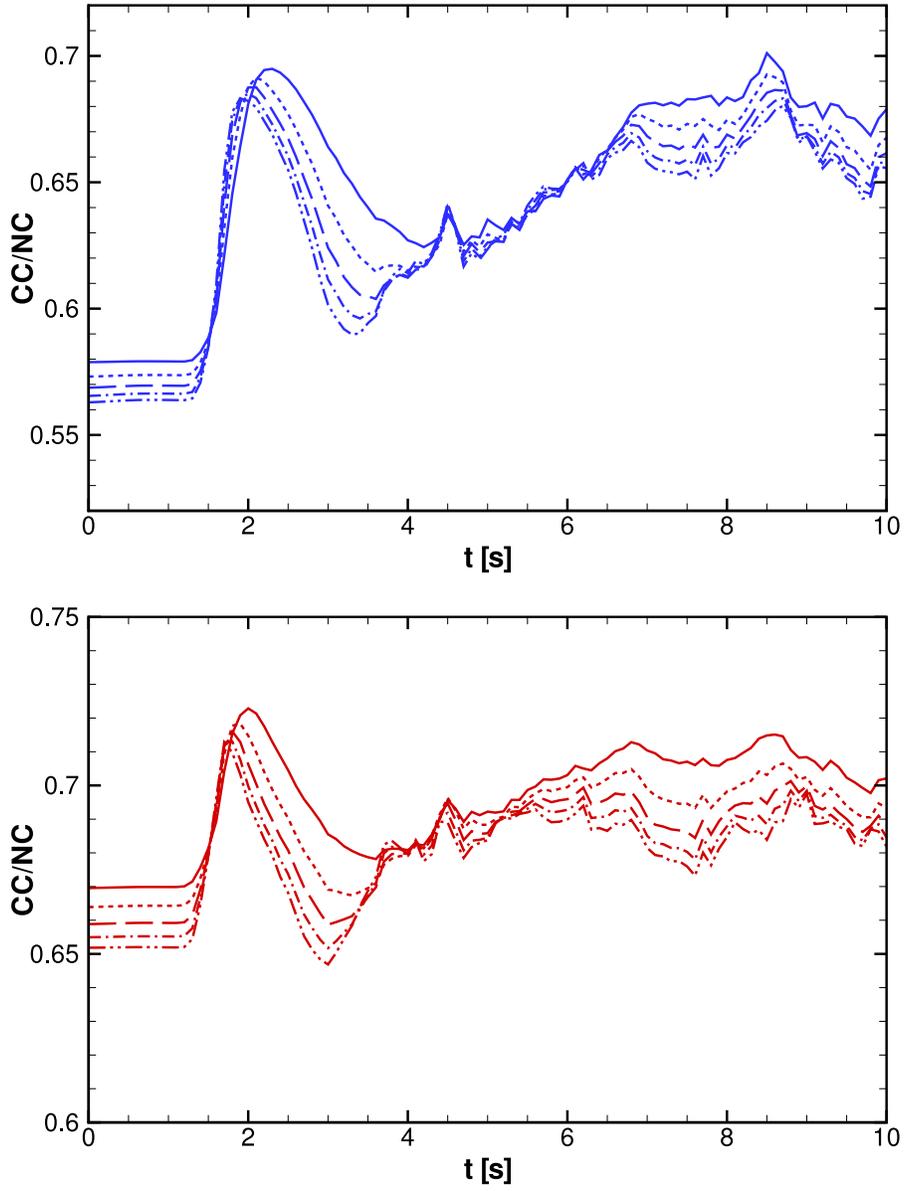}
\caption{\label{fig:ccnc.snv2d} The ratio of charged current event rates to neutral current event rates in 
a heavy water detector as a function of time for neutrino propagation through the $25^{\circ}$ 
radial slice of the the 2D simulation. The top panel is for an inverted hierarchy , the lower panel 
for a normal hierarchy. In each the lines represent different spectral parameters: 
$\alpha=1$ (solid), $\alpha=2$ (dotted), $\alpha=3$ (long dashed), 
$\alpha=4$ (dash dot) and $\alpha=5$ (long dash-double dot).}
\end{figure}
\begin{figure}
\includegraphics[width=5in]
{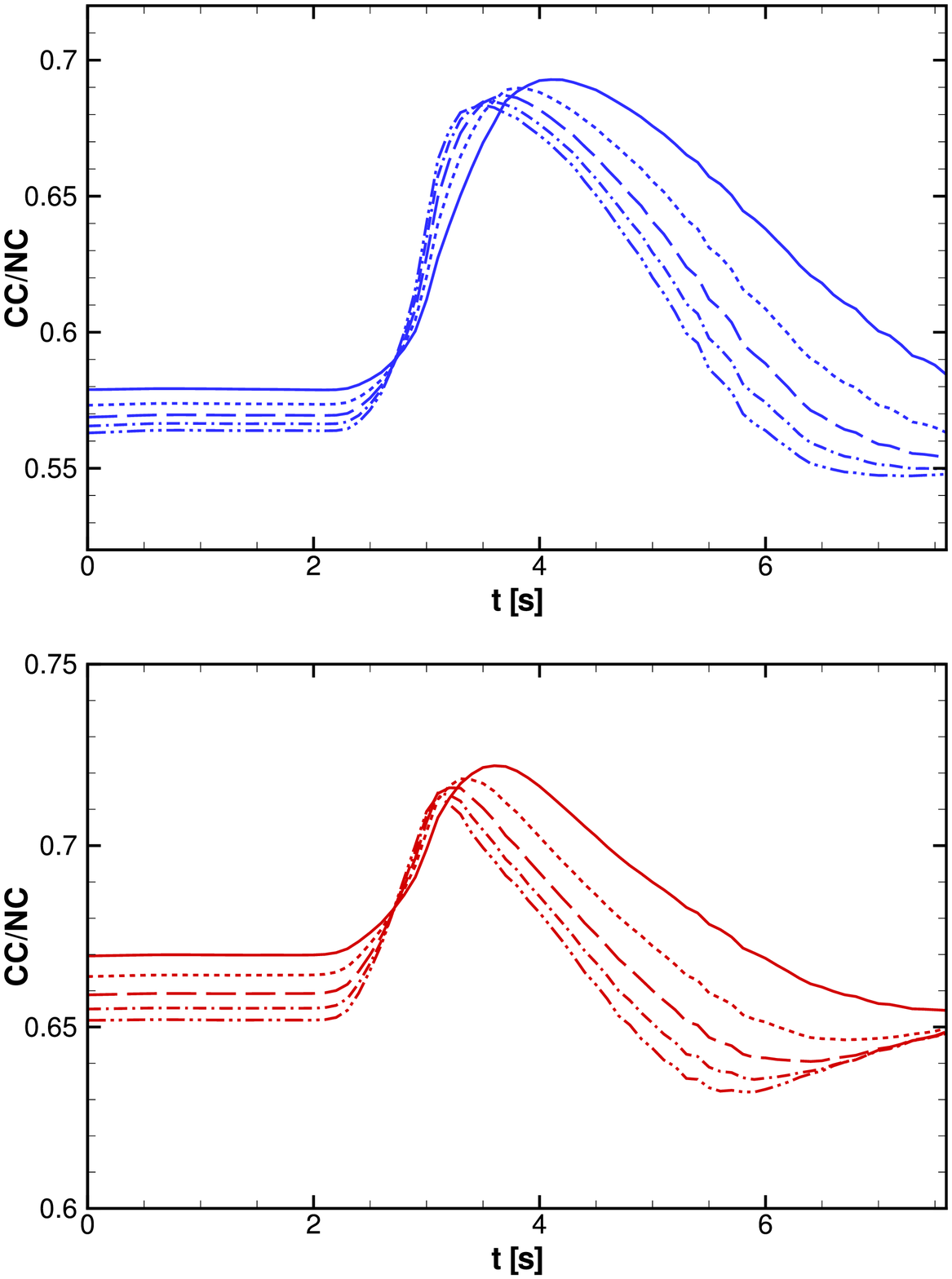}
\caption{\label{fig:ccnc.snv1d.1e35} The ratio of charged current event rates to neutral current event rates in 
a heavy water detector as a function of time for neutrino propagation through 
the 1D simulation with $Q=1.66 \times 10^{51}\;{\rm erg}$. The top panel is for an inverted hierarchy, the lower panel 
for a normal hierarchy. In each the lines are: $\alpha=1$ (solid), $\alpha=2$ (dotted), $\alpha=3$ (long dashed), 
$\alpha=4$ (dash dot) and $\alpha=5$ (long dash-double dot).}
\end{figure}
\begin{figure}
\includegraphics[width=5in]
{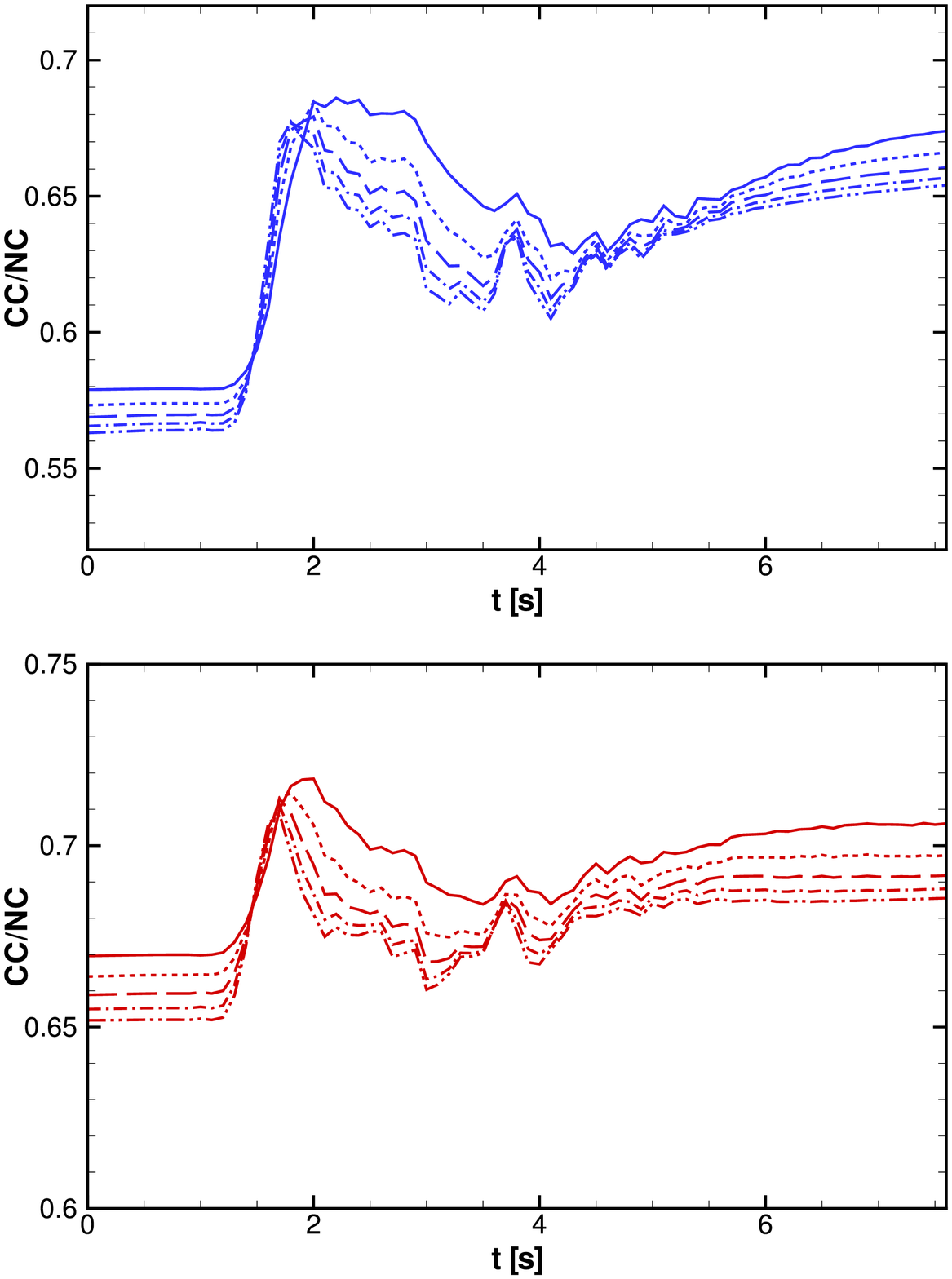}
\caption{\label{fig:ccnc.snv1d.8e35} The ratio of charged current event rates to neutral current event rates in 
a heavy water detector as a function of time for neutrino propagation through 
the 1D simulation with $Q=3.36 \times 10^{51}\;{\rm erg}$. The top panel is for 
an inverted hierarchy, the lower panel 
for a normal hierarchy. In each the lines are: $\alpha=1$ (solid), 
$\alpha=2$ (dotted), $\alpha=3$ (long dashed), 
$\alpha=4$ (dash dot) and $\alpha=5$ (long dash-double dot).}
\end{figure}
\begin{figure}
\includegraphics[width=5in]
{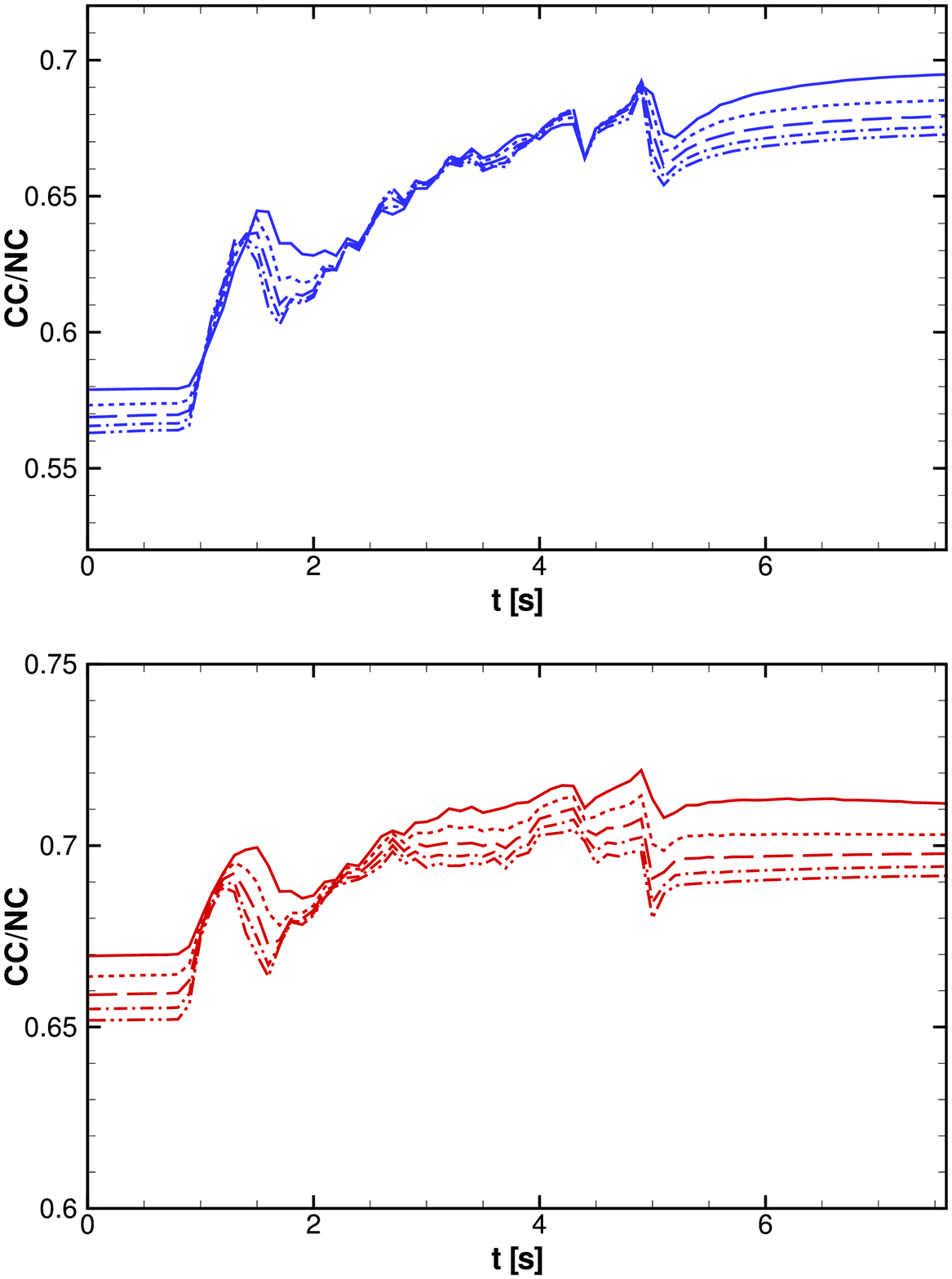}
\caption{\label{fig:ccnc.snv1d.2e36} The ratio of charged current event rates to neutral current events rates in 
a heavy water detector as a function of time for neutrino propagation through 
the 1D simulation with $Q=4.51 \times 10^{51}\;{\rm erg}$. The top panel is for an inverted hierarchy, the lower panel 
for a normal hierarchy. In each the lines are: $\alpha=1$ (solid), $\alpha=2$ (dotted), $\alpha=3$ (long dashed), 
$\alpha=4$ (dash dot) and $\alpha=5$ (long dash-double dot).}
\end{figure}
The first detector type we consider is a heavy water, SNO  
like, detector and the signal we focus upon is the ratio of charged current (CC) - both from electron neutrinos and 
electron antineutrinos - to neutral current (NC) event rates. 
This ratio of event rates is the same signal considered by Schirato \& Fuller \cite{SF2002} 
even though the $^{3}He$ counters and then the NaCl inserted into 
SNO gave the detector some capability of distinguishing between the $\nu_{e}$ and $\bar{\nu}_{e}$ events. 
To some extent the uncertainty of the initial spectra can be removed by considering this ratio of event 
rates though by no means is it eliminated. For the results that follow we calculate 
the event rates using the Nakamura \emph{et al.} \cite{2002NuPhA.707..561N} 
neutrino-deuteron cross sections. 

The basic understanding of what happens to the $\nu_{e}$ or $\bar{\nu}_{e}$ flux 
helps us understand the temporal evolution of the CC/NC ratio shown in Fig. (\ref{fig:ccnc.snv2d}) 
for the 2D simulation and for the 1D simulations in Figs. 
(\ref{fig:ccnc.snv1d.1e35}), (\ref{fig:ccnc.snv1d.8e35}) and (\ref{fig:ccnc.snv1d.2e36}). 
For all combinations of hierarchy, pinch parameter and explosion energy the
ratio begins to vary at around $1-2.5\;{\rm s}$ into the simulation. The initial increase 
in the ratio is due to larger charged current rates (the neural current rate is flavor blind) 
caused by greater electron neutrino flavor content (for the NH) or electron antineutrino flavor (for the IH). 
The change is a clear signature of a transition from adiabaticity to non-adiabaticity that occurs when 
the forward shock reaches the H resonances for neutrinos with $E \sim {\cal O}(10\;{\rm MeV})$. 
As expected, the ratio varies more strongly for an inverted hierarchy due to 
the greater change in the electron antineutrino spectrum when the profile becomes non-adiabatic 
than for the electron neutrinos as indicated by Figs. (\ref{fig:Fnu.alpha=3.IH.PHbar=0,1}) 
and (\ref{fig:Fnu.alpha=3.NH.PH=0,1}). More detailed examination of the figures reveals 
that for the particular case of the weak explosion - Fig. (\ref{fig:ccnc.snv1d.1e35}) - there is 
evidence of the shock effects moving up through the neutrino spectra because 
at late times we see the ratio dips beneath it's $t=0.0$ value. 

For the other simulations the CC/NC ratio exhibits a great deal of 
variability at the level of a few percent that continues 
for several seconds. We also notice that 
the variability of the $Q=3.36\times 10^{51}\;{\rm erg}$ and $Q=4.51\times 10^{51}\;{\rm erg}$ 
1D simulations ceased after $\sim 5-6\;{\rm s}$ when the forward shock had swept through the 
spectrum and the reverse shock had turned around while the CC/NC ratio for the 
2D simulation was sustained for the entire $10\;{\rm s}$. 
While observing a period of fluctuations in the CC/NC 
ratio would indicate that the explosion and the density profile was dynamic 
it is difficult to be more quantitative about the presence and motion of the various 
features of the profile from the use of this ratio of event rates alone.


\subsection{Water Cerenkov Detectors}

The second example of a detector we examine is a water Cerenkov type since this 
is the dominant supernova neutrino detection technology.
We focus upon the positron spectrum created by inverse $\beta$ reactions 
of antineutrinos upon protons. The temporal variations of the neutrino signal 
we have discussed in the paper would therefore require an inverted hierarchy in order to 
be visible in this signal. The positron spectrum seen in a detector, $\Phi_{e^+}(E_{e^+})$, 
is given by 
\begin{equation}
\Phi_{e^+} (E_{e^+}) = N_{p}\,\int\,dE_{\bar{\nu}_{e}}\,F_{\bar{\nu}_{e}}\,\frac{d\sigma}{dE_{e^+}}
\end{equation}
where $N_{p}$ is the number of protons in the detector, $F_{\bar{\nu}_{e}}$ is the electron antineutrino flux
and $d\sigma/dE_{e^+}$ is the differential cross section. For our analysis we used the 
cross sections of Strumia and Vissanti \cite{SV2003}. 
The positron spectrum 
allows us to see the effect of 
the density profile upon the neutrino flux as a function of neutrino energy but, before 
we show our results, we firstly consider the relationship between 
the neutrino energy and the positron energy and our ability to resolve the features of the 
neutrino spectrum. 
\begin{figure}
\includegraphics[width=5in]{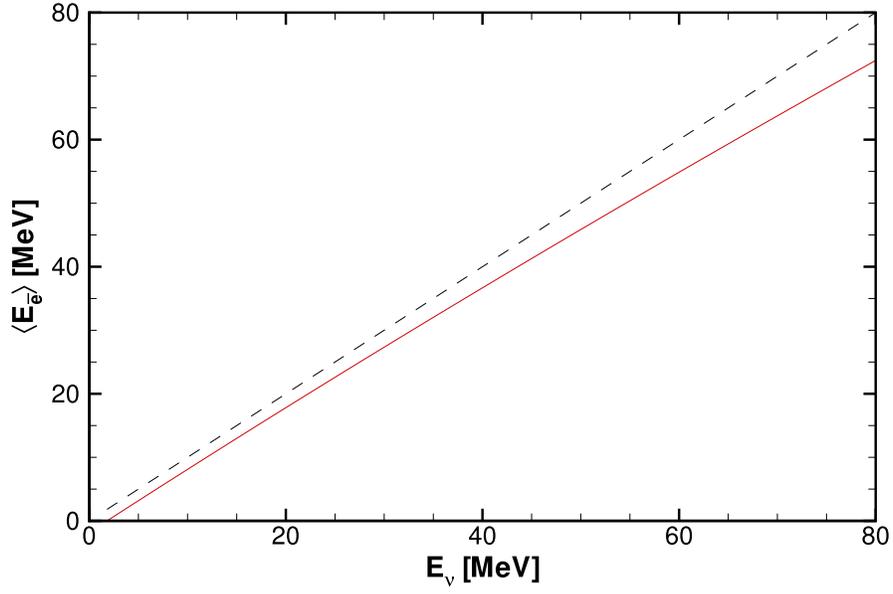}
\caption{\label{fig:meanEae.eps} The `mean' positron energy $\langle E_{\bar{e}}\rangle$ 
as a function of the neutrino energy (solid line) given by equation (16) in Strumia and Vissanti \cite{SV2003}. 
The dashed line is $\langle E_{\bar{e}}\rangle = E_{\nu}$.}
\end{figure}
The positron energy and neutrino energy are closely correlated as 
shown in Fig. (\ref{fig:meanEae.eps}). 
Features in the neutrino spectrum at a particular energy will appear in the positron spectrum at more or less the 
same energy. 
\begin{figure}
\includegraphics[width=5in]{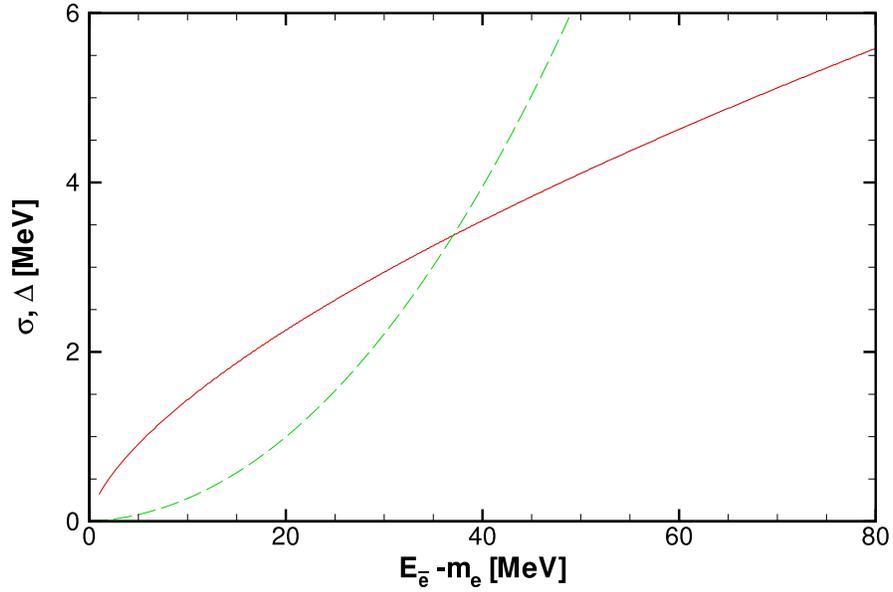}
\caption{\label{fig:SKerror,spread} The Super-K energy resolution error, $\sigma$, (solid line) and the 
range of positron energies, $\Delta$, (dashed) as a function of the positron kinetic energy.}
\end{figure}
But the positron spectrum is smoothed relative to the neutrino spectrum by two effects: 
the distribution of positron energies for a given neutrino energy, and the energy resolution of the 
detector. 
For a given neutrino energy the positrons emerge with a range of energies which we may invert 
so that for a given positron energy, $E_{e^+}$, there is a range of neutrino energies, $\Delta$, 
from which it could have arisen. This 
range\footnote{The formula is valid for positron energies such that $p_{e^+}+E_{e^+} \leq m_{p}$.} is given by 
\begin{equation}
\Delta =\frac{ m_{n}^{2}-m_{p}^{2}-m_{e}^{2} +2\,m_{p}\,E_{e^+}}{2}\,
\left( \frac{1}{m_{p}-p_{e^+}-E_{e^+}} - \frac{1}{m_{p}+p_{e^+}-E_{e^+}} \right)
\end{equation}
where $p_{e^+}$ is the momentum of the positron and $m_{n}$, $m_{p}$ and $m_{e}$ are the 
masses of the neutron, proton and electron respectively. 
$\Delta$ is shown in Fig. (\ref{fig:SKerror,spread}) as a function of the positron kinetic energy. 
And then for the energy resolution of the detector we used a filter so that the 
spectrum of positrons at energy $E_{e^+}$ is given by
\begin{equation}
\Phi(E_{e^+})=\frac{\int\,dE_{e^+}'\,\Phi(E_{e^+}')\,W(E_{e^+}|E_{e^+}')}{\int\,dE_{e^+}'\,W(E_{e^+}|E_{e^+}')}
\end{equation}
and we take $W(E_{e^+}|E_{e^+}')$ to be a normalized 
Gaussian with a mean $E_{e^+}'$ and a variance $\sigma^{2}(E_{e^+}')$ given 
by 
\begin{equation}
\frac{\sigma}{E_{e^+} -m_{e}}=0.319\,\left(\frac{1}{E_{e^+}-m_{e} [{\rm MeV}]}\right)^{0.3467}.
\end{equation}
This formula is our fit to the energy resolution of the Super-K detector \cite{Hosaka:2005um}. 
The energy resolution of the detector, $\sigma$, is also shown in figure (\ref{fig:SKerror,spread})
as a function of the positron kinetic energy. Both smearing effects, the range, $\Delta$, 
of neutrino energies 
that can produce a positron with energy $E_{e^+}$, and 
the detector resolution, $\sigma$, increase with $E_{e^+}$ and we notice that 
beyond $E_{e^+} \sim 40\;{\rm MeV}$ it is the range $\Delta$ that determines 
that determines the scale of the features we can resolve. Below 
$E_{e^+} \sim 40\;{\rm MeV}$ the detector resolution limits out ability to 
see features of the neutrino spectrum. Future detectors will improve upon the 
energy resolution used here and so - since this is also the energy range where we should focus our 
attention because it is where the electron antineutrino flux will vary the largest 
due to the changing adiabaticity - in all our figures we plot both the spectrum as it would be measured 
with the Super-K energy resolution and with a detector that has perfect resolution. 
Of course there is a third effect one needs to consider: the Poisson error due to 
the finite number of detected events. This does not smooth out features of the spectrum 
but rather make them difficult to detect. In what follows we shall ignore the 
Poisson statistics. 

\begin{figure}
\includegraphics[width=5in]
{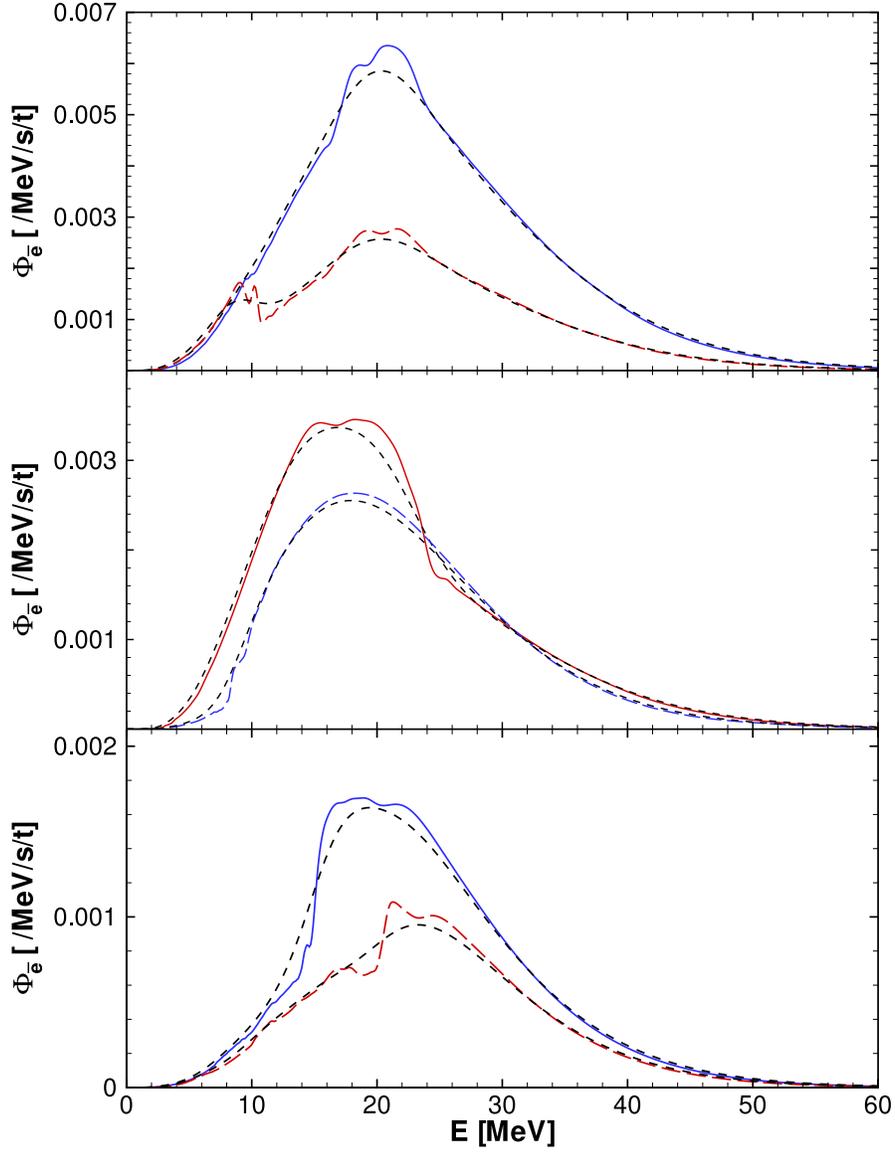} 
\caption{The positron spectrum per unit tonne of detector 
as a function of energy at various snapshot times 
for the 1D simulation with a total energy deposition of $Q=1.66\times 10^{51}\;{\rm erg}$. 
In the top panel the times are $t=0.0\;{\rm s}$ (solid) and $t=2.5\;{\rm s}$ (long dashed), in the middle panel 
$t=3.1\;{\rm s}$ (solid) and $t=4.0\;{\rm s}$ (long dashed), and in the bottom panel $t=5.0\;{\rm s}$ (solid) 
and $t=5.8\;{\rm s}$ (long dashed).
In all panels the short dashed lines indicate the spectrum convolved with our adopted detector energy 
resolution. \label{fig:FaevsE.1e35}}
\end{figure}
In Fig. (\ref{fig:FaevsE.1e35}) we plot the positron spectrum as a function of 
positron energy  in the case of the inverted hierarchy 
at various snapshot times using the neutrino flux we calculated 
for the weakest 1D simulation, $Q=1.66\times 10^{51}\;{\rm erg}$. With this spectral
information,  
we can begin to see the features of the explosion. The shock is clearly seen 
at $E\sim 10\;{\rm MeV}$ at $t=2.5\;{\rm s}$ in the perfectly resolved spectrum. 
By $t=3.1\;{\rm s}$ the sharp 
change in spectrum has moved up to $E\sim 25\;{\rm MeV}$ and by $t=4.0\;{\rm s}$ we begin to see 
the backside of the shock at $E\sim 10\;{\rm MeV}$ and again at $t=5.0\;{\rm s}$ at $E\sim 15\;{\rm MeV}$ 
and $t=5.8\;{\rm s}$ at $E\sim 20\;{\rm MeV}$.
The shock is a less dramatic feature in the spectra convolved with our adopted 
detector energy resolution but is visible particularly at $t=2.5\;{\rm s}$. 
Given enough events even present detector technology would allow us to see the shock 
move up through the positron spectrum from which we could begin to glean information about the 
explosion. 
\begin{figure}
\includegraphics[width=5in]
{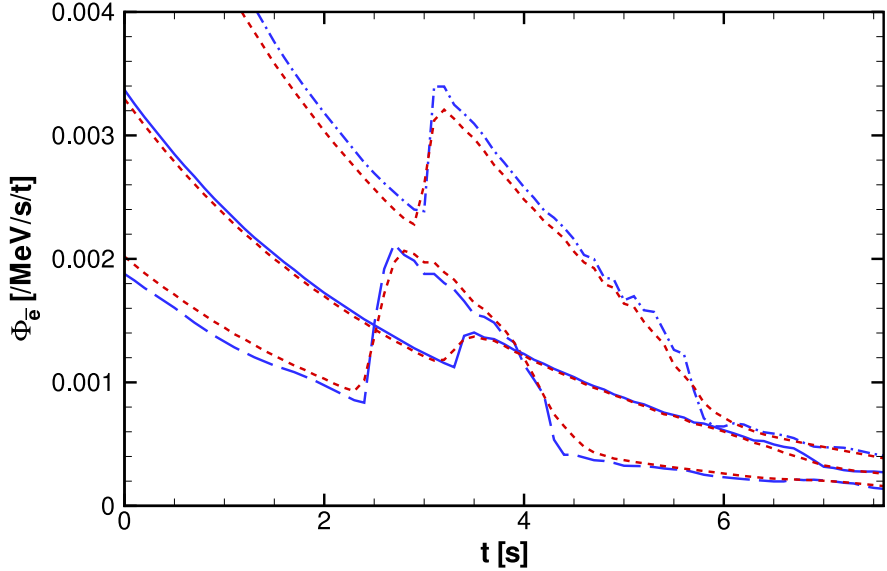} 
\caption{The positron spectrum in the case of the inverted hierarchy 
per unit tonne of detector as a function of time at various positron energies 
for the 1D simulation with a total energy deposition of $Q=1.66\times 10^{51}\;{\rm erg}$. 
The curves are $E=10\;{\rm MeV}$ (long dashed), $E=20\;{\rm MeV}$ (dash dot), and $E=30\;{\rm MeV}$ (solid).
The short dashed lines indicate the spectrum convolved with our adopted detector energy 
resolution.
\label{fig:Faevst.1e35}}
\end{figure}
The explosion features also appear in the evolution of the positron spectrum at particular energies 
as a function of time. This is shown in figure (\ref{fig:Faevst.1e35}) where we plot the 
positron spectrum for $E=10\;{\rm MeV}$, $E=20\;{\rm MeV}$ and $E=30\;{\rm MeV}$. 
The general trend for each energy is $\propto \exp(-t/\tau)$ since this was the time dependence we 
used for the emitted spectra but on top of this exponential decay the shock 
is obviously visible for all three chosen energies in both filtered and unfiltered curves. 
The width of the forward shock feature grows with $E$ as we noted earlier. 
\begin{figure}
\includegraphics[width=5in]
{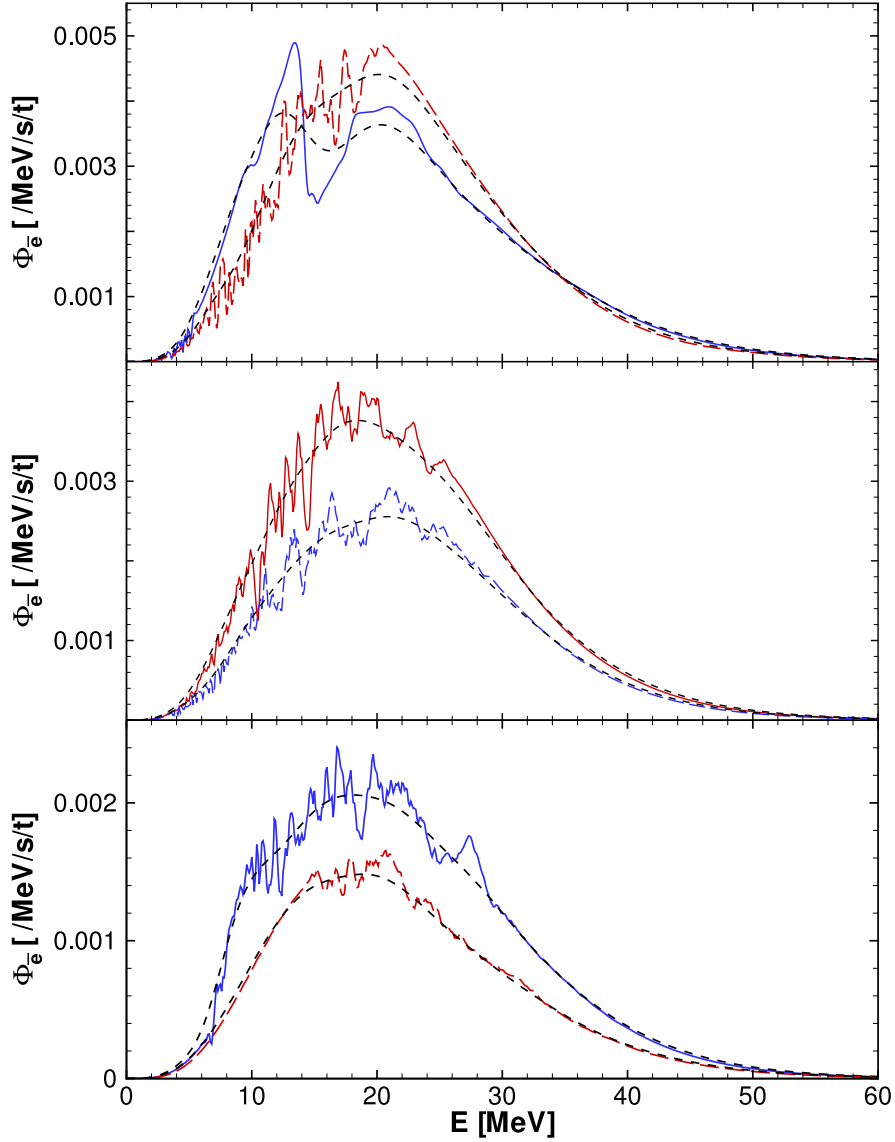} 
\caption{The positron spectrum in the case of the inverted hierarchy 
per unit tonne of detector as a function of time at various positron energies 
for the 1D simulation with a total energy deposition of $Q=3.36\times 10^{51}\;{\rm erg}$. 
In the top panel the times are $t=1.5\;{\rm s}$ (solid) and $t=2.1\;{\rm s}$ (long dashed), 
in the middle panel 
$t=2.4\;{\rm s}$ (solid) and $t=3.1\;{\rm s}$ (long dashed), and in the bottom panel 
$t=3.6\;{\rm s}$ (solid) and $t=4.9\;{\rm s}$ (long dashed).
In all panels the short dashed lines indicate the spectrum convolved with our adopted detector energy 
resolution.\label{fig:FaevsE.8e35}}
\end{figure}
In Fig. (\ref{fig:FaevsE.8e35}) we plot the positron spectrum for the more powerful explosion 
with $Q=3.36\times 10^{51}\;{\rm erg}$ at various snapshot times. The phase effects 
we saw in the crossing probability for this model appear also in the spectrum for perfect 
detector resolution. They are largest in the range of $10-30\;{\rm MeV}$ where the difference 
between adiabatic and non-adiabatic electron antineutrino flux is largest and where 
the distribution of positron energies for a given neutrino energy is narrow. At higher energies 
the phase effects largely disappear because of the smoothing effect of the positron distribution. 
\begin{figure}
\includegraphics[width=5in]
{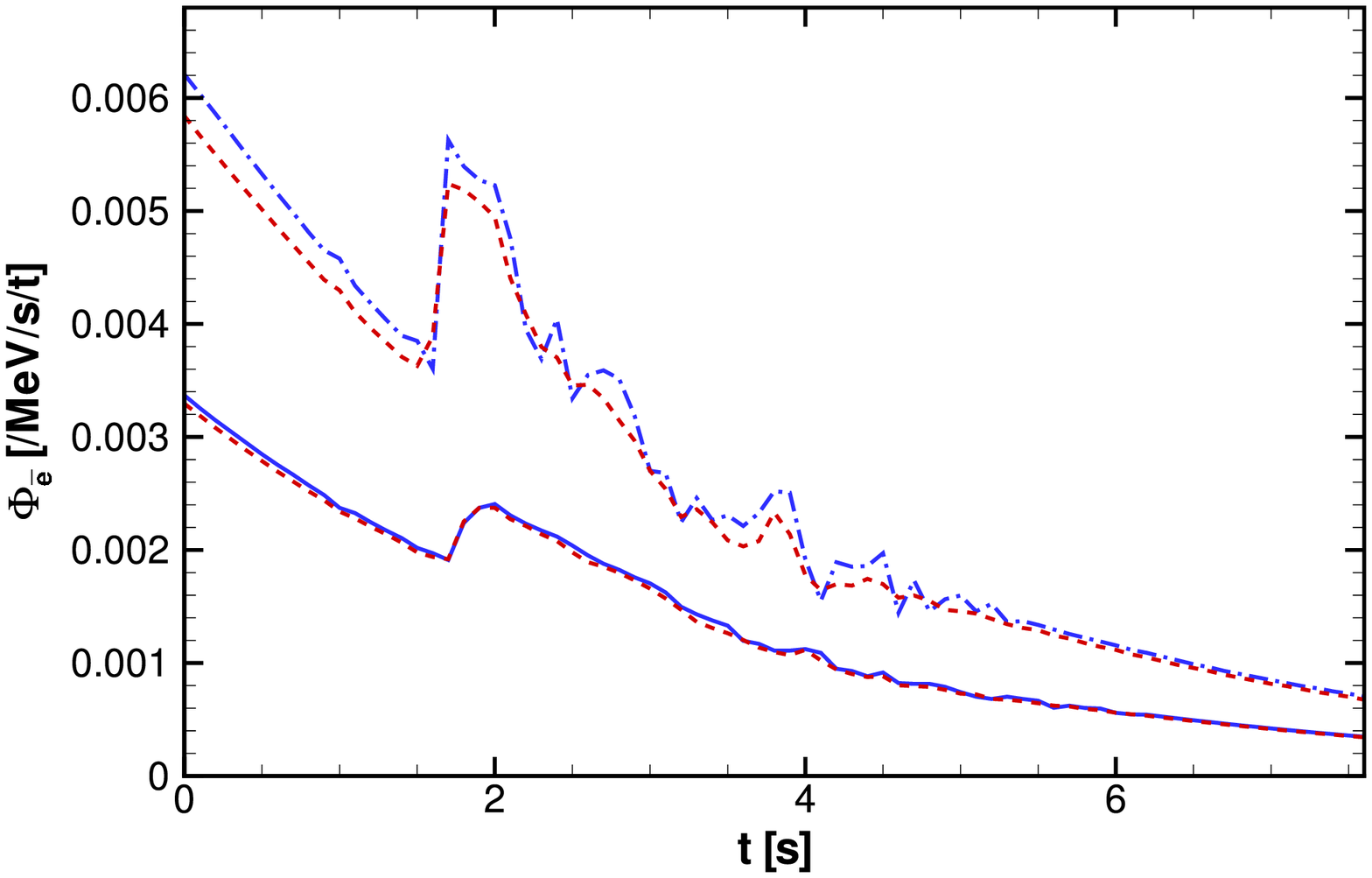} 
\caption{The positron spectrum in the case of the inverted hierarchy 
per unit tonne of detector as a function of time at various positron energies 
for the 1D simulation with a total energy deposition of $Q=3.36\times 10^{51}\;{\rm erg}$, and
an inverted hierarchy. 
The curves are $E=20\;{\rm MeV}$ (dash dot), and $E=30\;{\rm MeV}$ (solid).
The short dashed lines indicate the spectrum convolved with our adopted detector energy 
resolution.\label{fig:Faevst.8e35}}
\end{figure}
This smoothing is also seen in the plot of the spectrum at various energies as a function of time shown in Fig.  
(\ref{fig:Faevst.8e35}) since the $20\;{\rm MeV}$ positrons has much greater variability 
than then $30\;{\rm MeV}$ positrons even in the filtered spectrum. 
\begin{figure}
\includegraphics[width=5in]
{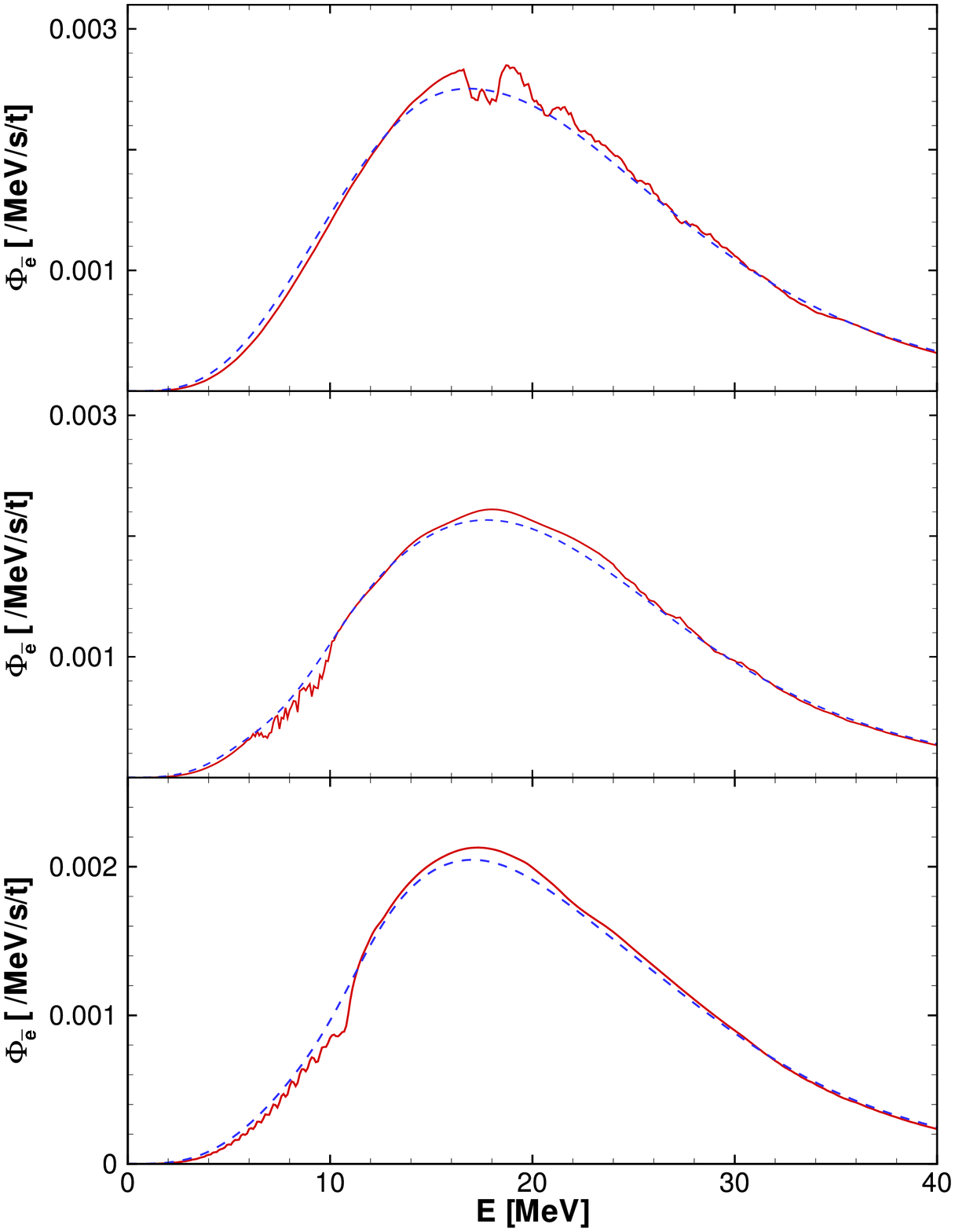} 
\caption{The positron spectrum per unit tonne of water Cerenkov detector 
as a function of energy at various snapshot times 
for the 1D simulation with a total energy deposition of $Q=4.51\times 10^{51}\;{\rm erg}$. 
In the top panel the time is $t=4.0\;{\rm s}$, in the middle panel 
$t=4.5\;{\rm s}$, and in the bottom panel $t=4.9\;{\rm s}$.  
In all panels the short dashed lines indicate the spectrum convolved with our adopted detector energy 
resolution.\label{fig:FaevsE.2e36}}
\end{figure}
We can even find the signature of a reverse shock that stalled and headed back to the core. 
In Fig. (\ref{fig:FaevsE.2e36}) we plot the positron spectrum for the more energetic 
$Q=4.51\times 10^{51}\;{\rm erg}$ 1D explosion. The snapshot times are the same 
as those shown in Fig. (\ref{fig:rho.snv1d.2e36.t=4,4.5,4.9}) for the density and 
Fig. (\ref{fig:PC.snv1d.2e36.t=4,4.5,4.9}) for the crossing probability for this model.
The returning reverse shock appears at $\sim 10\;{\rm MeV}$ at $t=4.5\;{\rm s}$ which 
matches what we expect from the crossing probability seen in the middle panel of 
Fig. (\ref{fig:PC.snv1d.2e36.t=4,4.5,4.9}). 

\begin{figure}
\includegraphics[width=5in]
{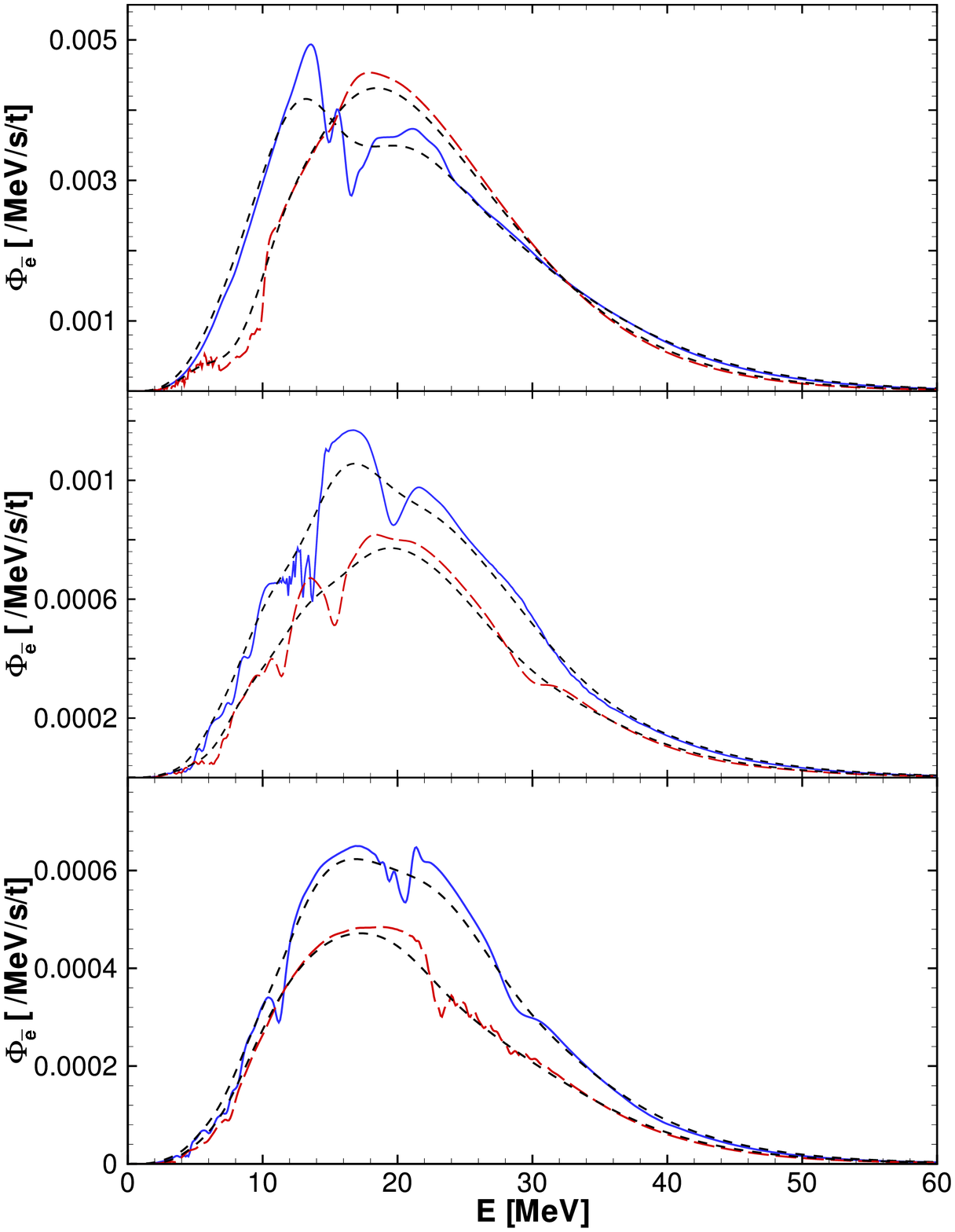} 
\caption{The positron spectrum per unit tonne of water Cerenkov detector 
as a function of energy at various snapshot times 
for the 2D simulation. 
In the top panel the times are $t=1.6\;{\rm s}$ (solid) and $t=2.4\;{\rm s}$ (long dashed), in the middle panel 
$t=6.4\;{\rm s}$ (solid) and $t=7.4\;{\rm s}$ (long dashed), and in the bottom panel $t=8.0\;{\rm s}$ (solid) 
and $t=9.0\;{\rm s}$ (long dashed).
In all panels the short dashed lines indicate the spectrum convolved with our adopted detector energy 
resolution.\label{fig:FaevsE.snv2d}}
\end{figure}
Finally in Fig. (\ref{fig:FaevsE.snv2d}) we plot the spectrum for the 2D simulation at various 
snapshot times selected so that they match as closely as possible the  snapshots we have shown of the 
profile and/or the crossing probability. The forward shock and phase effects can be seen 
here too in the unfiltered spectrum and the inward moving reverse shock is responsible for the
phase effects seen at $E\sim 15\;{\rm MeV}$ at $t=6.4\;{\rm s}$ and in Fig. 
(\ref{fig:PC.snv2d.t=2.4,5.4,6.4}). Other features upon the positron 
spectrum can also be related to various features shown in the crossing probability for this model.


\section{Summary and Conclusions}

The next Galactic supernova has the potential to provide a great deal of information about both 
neutrino mixing and about the supernova. The temporal and spectral evolution of the signal is 
altered as the star explodes due to the changes in the density profile of the star. 
The forward shock should be easily detected even with present detector technology 
as either a feature sweeping up through the positron spectrum in a water Cerenkov detector or as a 
sudden increase in event rates for a detector with no spectral information such as a heavy water detector. 
Once identified, and if the progenitor were known, the information can be used to measure the 
shock speed, the strength of the shock and potentially also the stalled shock time delay. 
The reverse shock can be spotted qualitatively as the cause of fluctuations in either 
event rates or upon the positron spectrum. There is also the possibility 
for the reverse shock to be seen if its motion is reversed 
and it heads back to the core. 
The neutrino signal may also contain hints that 
the supernova was aspherical as a qualitative change in the 
phase effects.  

\acknowledgments

\vspace{1.cm} The authors are very grateful to John Blondin for his helpful advice with 
the hydrodynamic simulations. 
This work was supported by the U.S. Department of 
Energy at UMn under grant DE-FG02-87ER40328 and at 
NCSU under grant DE-FG02-02ER41216..


\end{document}